\documentclass[%
aip,
jcp,
amsmath,amssymb,
preprint,%
]{revtex4-1}

\usepackage{graphicx}
\usepackage{dcolumn}
\usepackage{bm}

\usepackage[utf8]{inputenc}
\usepackage[T1]{fontenc}
\usepackage{mathptmx}
\usepackage{etoolbox}

\usepackage{comment}

\usepackage{enumerate}
\usepackage{algorithm}
\usepackage{algorithmicx}
\usepackage{algpseudocode}
\usepackage{paralist}
\usepackage{threeparttable}
\usepackage{threeparttablex}
\usepackage{booktabs}
\usepackage{makecell}
\usepackage{multirow}
\usepackage{color}

\algnewcommand{\Inputs}[1]{
  \State \textbf{Inputs:}
  \Statex \hspace*{\algorithmicindent}\parbox[t]{.8\linewidth}{\raggedright #1}
}
\algnewcommand{\Initialize}[1]{
  \State \textbf{Initialize:}
  \Statex \hspace*{\algorithmicindent}\parbox[t]{.8\linewidth}{\raggedright #1}
}
\algnewcommand{\Outputs}[1]{
  \State \textbf{Outputs:}
  \Statex \hspace*{\algorithmicindent}\parbox[t]{.8\linewidth}{\raggedright #1}
}
\algnewcommand\algorithmicswitch{\textbf{switch}}
\algnewcommand\algorithmiccase{\textbf{case}}
\algdef{SE}[SWITCH]{Switch}{EndSwitch}[1]{\algorithmicswitch\ #1\ \algorithmicdo}{\algorithmicend\ \algorithmicswitch}%
\algdef{SE}[CASE]{Case}{EndCase}[1]{\algorithmiccase\ #1}{\algorithmicend\ \algorithmiccase}%
\algtext*{EndSwitch}%
\algtext*{EndCase}%

\newfloat{Algorithm}{htbp}{alg}
\floatname{Algorithm}{Algorithm}

\makeatletter
\def\@email#1#2{%
 \endgroup
 \patchcmd{\titleblock@produce}
  {\frontmatter@RRAPformat}
  {\frontmatter@RRAPformat{\produce@RRAP{*#1\href{mailto:#2}{#2}}}\frontmatter@RRAPformat}
  {}{}
}%
\makeatother

\UseRawInputEncoding
\begin{document}

\preprint{AIP/123-QED}

\title[Self-Adaptive Real-Time Time-Dependent Density Functional Theory for X-ray Absorptions]{Self-Adaptive Real-Time Time-Dependent Density Functional Theory for X-ray Absorptions}
\author{Linfeng Ye}
\affiliation{
Qingdao Institute for Theoretical and Computational Sciences, Institute of Frontier and Interdisciplinary Science, Shandong University, Qingdao, Shandong 266237, P. R. China}

\author{Hao Wang*}\email{wanghaosd@sdu.edu.cn} %
\affiliation{
Qingdao Institute for Theoretical and Computational Sciences, Institute of Frontier and Interdisciplinary Science, Shandong University, Qingdao, Shandong 266237, P. R. China}

\author{Yong Zhang}
\affiliation{
Qingdao Institute for Theoretical and Computational Sciences, Institute of Frontier and Interdisciplinary Science, Shandong University, Qingdao, Shandong 266237, P. R. China}

\author{Wenjian Liu*}\email{liuwj@sdu.edu.cn}%
\affiliation{
Qingdao Institute for Theoretical and Computational Sciences, Institute of Frontier and Interdisciplinary Science, Shandong University, Qingdao, Shandong 266237, P. R. China}


\begin{abstract}
Real-time time-dependent density functional theory (RT-TDDFT) can in principle access the whole absorption spectrum of a many-electron system exposed to a narrow pulse.
However, this requires an accurate and efficient propagator for the numerical integration of the time-dependent Kohn-Sham equation.
While a low-order time propagator is already sufficient for the low-lying valence absorption spectra, it is no longer the case for the X-ray absorption
spectra (XAS) of systems composed even only of light elements, for which the use of a high-order propagator is indispensable.
It is then crucial to choose a largest possible time step and a shortest possible simulation time, so as to minimize the computational cost.
To this end, we propose here a robust AutoPST approach to determine automatically (Auto) the propagator (P), step (S), and time (T)
for relativistic RT-TDDFT simulations of XAS.
\end{abstract}

\maketitle

\onecolumngrid

\section{Introduction}
Triggered by the development of advanced synchrotron light sources and X-ray free electron lasers,
X-ray absorption spectroscopy (XAS) has become an increasingly important technique for
probing local electronic and geometric structure of matter. In XAS, innermost core electrons are
excited to bound valence or continuum states in energetically distinct absorption edges, which
are conventionally labeled according to the origins of the electronic transitions ($K$-edge for $1s$, $L_1$-edge for $2s$, $L_2$-edge for $2p_{1/2}$,
and $L_3$-edge for $2p_{3/2}$, etc). There are two main regions in an XAS spectrum,
X-ray absorption near-edge structure (XANES) near the rising edge with sharp resonance peaks
and extended X-ray absorption fine structure (EXAFS) after the XANES region with gentle oscillations.
The former provides electronic fingerprints of a particular atom (e.g., oxidation state, coordination, and bonding, etc), whereas the latter
gives structural information about neighboring atoms (e.g., identity, distances, and coordination/solvation shells, etc).
While the essential quantities (scattering amplitudes and phase shifts)
in EXAFS can well be modeled already by a simple, damped spherical photoelectron wave function approximation,
the complicated features of XANES spectra require more sophisticated treatments\cite{XASChemRev2018}, among which
both static and time-dependent density functional theory (TDDFT) are widely adopted due to their good tradeoff between
accuracy and efficiency. The former is commonly referred to as delta self-consistent field ($\Delta$SCF) and has a very appealing feature:
orbital relaxation (which is of vast importance for core excitations) is fully accounted for by SCF calculation of each
core excited configuration. In particular, full spin adaptation can be achieved within the framework of multi-state DFT\cite{MSDFT2016,MSDFT2017,MSDFT2019,coreMSDFT}.
Yet, such one-by-one calculations of core excited states may not always be possible due to the non-Aufbau nature, given the availability of
many algorithms\cite{MOM2008,mom2014,CoreDFT2015,CoreDFT2017prl,CoreDFT2017,CoreDFT2018,CoreDFT2020,CoreDFT2020a,CoreDFT2020b,CoreDFT2020c,CoreDFT2020d}.
In contrast, TDDFT can access all core excited states in one shot, by either diagonalization as in
linear response (LR) TDDFT\cite{CoreLR-TD2004,CoreLR-TD2003,CoreLR-TD2007,CoreLR-TD2010,CoreLR-TD2011,CoreLR-TD2012,SARPA,SATDDFT,SATDDFTx,CoreLR-TD2013,CoreDFT2016,CoreLR-TD2019,CoreLR-TD2019a} or spectral analysis of the time signal generated by real-time (RT) TDDFT
\cite{CoreRT-TD2012,CoreRT-TD2013,CoreRT-TD2015,CoreRT-TD2016,PadeFitting2016,CoreRT-TD2019,CoreRT-TD2020,LiXSChemRev2020,IntruderRT-TD2022}.
Yet, both LR-TDDFT and RT-TDDFT lacks orbital relaxation, thereby yielding XANES spectra that usually have large shifts relative to the experimental ones.
Since core excitations look very much like charge transfer excitations, it is clear that such large shifts can be alleviated to a large extent by using
tuned range-separated functionals\cite{CoreRT-TD2015,CoreTD-IP2016,LR-TD2019RS}. Anyhow, peak separations predicted by LR-TDDFT and RT-TDDFT
are usually very good. As such, such one-for-all approaches are much more appealing than the one-by-one $\Delta$SCF approaches,
at least from the computational point of view. Naturally, the relativistic counterparts of LR-TDDFT\cite{BDFTD1,BDFTD2,Saue4CTD,NonColl2019,4CTDDFLRT2019,2CZORATDDFT,BDFTD3,X2CTDDFT1,X2CTDDFT2,2CECPTDDFT,Weigend2014,2CZORATD2014,LiX2CTD2016,LiX2CTD2017,X2CLR-TD2019,lwjTDRev} or RT-TDDFT\cite{RT4CTD2015,RT4CTD2018,RT4CTD2020,RTX2CTD1,RTX2CTD2,RTX2CTD3,LiXSChemPhysRev2020,darapaneni2020simulated} should be invoked
to account for relativistic effects, which are sizable already for core excitations even of very light elements.
As for nonrelativistic/relativistic LR-TDDFT and RT-TDDFT themselves, LR-TDDFT is certainly advantageous over RT-TDDFT in accessing dark states
(which are often important for, e.g., excited energy/charge transfers) as well as
in the assignment of excited states due to the availability of CIS (configuration interaction singles) type of state eigenvectors\cite{Casida1995},
especially when full molecular symmetry is employed\cite{Symm2009}. Moreover, there exist very robust algorithms\cite{iVI,iVI-TDDFT,CSFD2019}
that can access directly the core excited states as interior roots of the LR-TDDFT eigenvalue problem,
without the need to invoke the so-called core-valence separation\cite{CoreLR-TD2003,CoreLR-TD2007,CoreLR-TD2010,CoreLR-TD2011,CoreLR-TD2012},
a very good approximation though\cite{CVSgood}.
Nevertheless, RT-TDDFT has the following advantages over LR-TDDFT:
(1) only the exchange-correlation (XC) potential is needed, requiring no response of the XC potential,
(2) the entire absorption spectrum can be obtained at once, via a single spectral analysis of the time-dependent dipole moment, (3)
the memory footprint is only about one/two times that of the relativistic/nonrelativistic ground-state DFT calculation, thereby avoiding the memory bottleneck of
LR-TDDFT resulting from the use of a large number of trial vectors during the iterative partial diagonalization,
and (4) RT-TDDFT has the capability of accessing dynamical properties beyond the linear response regime.
Because of these, RT-TDDFT has gained great popularity in the last decades.

From the computational point of view, the very first issue of RT-TDDFT is how to design a suitable propagator for
the numerical integration of the underlying time-dependent Kohn-Sham equation (TDKS).
While a low-order propagator\cite{kosloff1988time,PropagatorsKS2004,PropagatorsKS2018} (e.g.,
the second-order Magus propagator (MP2)\cite{Magnus1954}) is already sufficient for the low-lying valence absorption spectra,
it is no longer the case for the XAS of systems composed even only of light elements: a low-order propagator requires an exceedingly small time step
to achieve a sufficient accuracy. The simulation is then much less efficient than the use of a higher-order propagator
along with a larger time step. It is found here that, for the same spectral accuracy ($<0.05$ eV),
the fourth-order optimized commutator-free exponential time-propagator (oCFET4)\cite{oCFET2011}
is most efficient among the tested high-order propagators (including MPn\cite{Magnus1954,MagnusRev}, CFETn\cite{CFET2006}, and oCFETn\cite{oCFET2011} with n=4, 6)
and can hence be regarded as a `just good enough' high-order time propagator for simulations of XAS.
However, this is only true when it is combined with the EPEP3 (exponential prediction of density matrix
and exponential correction of density matrix at three Gauss-Legendre points) type of predictor-corrector proposed here.
Moreover, very robust linear relations and block aliasing conditions (BAC) for automatic determination of the largest possible time step
as well as a good estimate of the shortest possible total simulation time will be established to automate the EPEP3@oCFET4/RT-TDDFT simulations
of XAS. Because of the automated nature, the algorithms will be dubbed collectively as AutoPST (automated determination of propagator, step, and time),
which can further be combined with full molecular symmetry\cite{Symm2009}.
To make the presentation complete, we first recapitulate the TDKS in Sec. \ref{TDKS}
and then discuss the implementation of oCFET4 in Sec. \ref{SecPropagator} and EPEP3 in Sec. \ref{SecPC}, where the closely related CFET4\cite{CFET2006} and EPEP2
are also presented for comparison.
The automated determinations of time step and total time are detailed
in Secs. \ref{SecStep} and \ref{SecTotT}, respectively. Illustrative examples are then provided in Sec. \ref{Results} to reveal
the performance of the proposed AutoPST
approach. Some concluding remarks are finally made in Sec. \ref{Conclusion}.

\section{Integration of TDKS}\label{Theory}
\subsection{TDKS}\label{TDKS}
RT-TDDFT amounts to integrating numerically the TDKS equation, followed by a spectral analysis to extract information on the electronic transitions.
Under the adiabatic approximation, the TDKS equation discretized in a time-independent, orthonormal basis $\{\chi_{\mu}\}$ can be written as
\begin{equation}
  i \frac{d}{dt}\mathbf{P} (t) = [\mathbf{F}(t),\mathbf{P} (t)],\quad \mathbf{F}(t)= \mathbf{F}_0(t)+ \mathbf{V}^{(1)}(t), \label{LvN}
\end{equation}
which is also called Liouville-von Neumann equation for the evolution of the one-particle density matrix (1PDM) $\mathbf{P}(t)$.
Here, $\mathbf{F}_0(t)$ is the matrix representation of the field-free KS Hamiltonian at time $t$,
\begin{align}
 F_0(t)&= T + V_{\mathrm{nuc}}+ V_h [\rho(t)] + c_{\mathrm{x}}\Sigma_{\mathrm{x}}[\gamma(t)]+c_{\mathrm{xc}}V_{\mathrm{xc}} [\rho(t)],\label{F0op}\\
 \rho(\boldsymbol{r}, t)&= \sum_{\mu\nu} \chi_{\mu}^\dag(\boldsymbol{r})\chi_{\nu}(\boldsymbol{r}) P_{\nu\mu}(t),\nonumber\\
  \gamma(\boldsymbol{r}, \boldsymbol{r}', t)&=\sum_{\mu\nu} \chi_{\nu}(\boldsymbol{r})\chi_{\mu}^\dag(\boldsymbol{r}') P_{\nu\mu}(t),\label{Den}
\end{align}
which is composed of the kinetic energy operator $T$ (nonrelativistic or relativistic\cite{LiuPerspective2020,LiuSciChina2020}),
(static) nuclear attraction $V_{\mathrm{nuc}}$, Hartree potential $V_h$, exact exchange potential $\Sigma_{\mathrm{x}}$, and
XC potential $V_{\mathrm{xc}}$. The scaling coefficients $c_{\mathrm{x}}$ and $c_{\mathrm{xc}}$ denote the respective portions of $\Sigma_{\mathrm{x}}$
and $V_{\mathrm{xc}}$ included in the chosen hybrid functional. As for the time-dependent external field, we consider here simply
a delta-type impulse
 \begin{equation}
 \boldsymbol \varepsilon(\boldsymbol{r},t)=\kappa\boldsymbol{n}  \delta(t),
\end{equation}
where $\kappa$ and $\boldsymbol{n}$ represent the field strength and polarization unit vector, respectively.
The system-light interaction operator $\mathbf{V}^{(1)}(t)$ in Eq. \eqref{LvN} then reads (under the dipole approximation)
\begin{equation}
\mathbf{V}^{(1)}(t)=\tilde{\mathbf{V}}^{(1)}\delta(t),\quad \tilde{V}^{(1)}_{\mu\nu}= \kappa
\sum_{\alpha=x,y,z} \langle \chi_{\mu}|r_{\alpha}|\chi_{\nu}\rangle n_{\alpha}.\label{DipMat}
\end{equation}

Given an initial value $\mathbf{P}(t_0)$, the solution of Eq. \eqref{LvN} can formally be written as
\begin{equation}
  \mathbf{P} (t) = \mathbf{U} (t, t_0) \mathbf{P}(t_0)\mathbf{U}^{\dagger} (t, t_0),
\end{equation}
where $\mathbf{U}(t, t_0)$ takes the following form
\begin{align}
  \mathbf{U} (t, t_0) & =  \mathcal{T}\mathrm{exp}\{- i \int_{t_0}^t \mathbf{F}[\tau] d\tau\} \label{Uop}\\
  &=\sum_{n=0}^{\infty}\frac{(-i)^n}{n!}\int_{t_0}^td\tau_1\int_{t_0}^t d\tau_2\cdots \int_{t_0}^t d\tau_n\mathcal{T}[\mathbf{F}(\tau_1)\mathbf{F}(\tau_2)\cdots\mathbf{F}(\tau_n)].
\end{align}
Here, $\mathcal{T}$ is the time-ordering operator accounting for the fact that the KS matrices at different times do not commute.
In practice, the time interval $t-t_0$ will be split into $N$ slices with equal length $\Delta t$, which amounts to splitting $\mathbf{U}(t, t_0)$ as
\begin{equation}
 \mathbf{U} (t, t_0) = \prod^{N - 1}_{k = 0}\mathbf{U} (t_{k + 1}, t_k),\quad t_k=t_0+k \Delta t.\label{Uproduct}
\end{equation}
In particular, $\mathbf{U}(0^+,0^-)$ takes a very simple for the delta-perturbation \eqref{DipMat},
\begin{equation}
\mathbf{U}(0^+,0^-)=\mathrm{exp}(-i\tilde{\mathbf{V}}^{(1)}),\label{U00mat}
\end{equation}
which can readily be obtained from Eq. \eqref{Uop} by noticing that $\mathbf{F}_0(\tau)$
does not change during the infinitely small time interval $[0^-,0^+]$, such that only the $\delta(t)$ term contributes to the time integral
(see Ref. \citenum{RT4CTD2015} for a more rigorous derivation). This suggests that
a perturb-then-propagate procedure\cite{rt-LR-TDDFT2004,Wang2007,RT4CTD2015} can be invoked for the evolution of $\mathbf{P}(t)$, viz.,
\begin{align}
\mathbf{P} (t) &= \mathbf{U}(t,0^+)\mathbf{P} (0^+)\mathbf{U}^{\dagger}(t,0^+),\\
\mathbf{P} (0^+)&=\mathrm{exp}(-i\tilde{\mathbf{V}}^{(1)})\mathbf{P} (0^-)\mathrm{exp}(i\tilde{\mathbf{V}}^{(1)}),\label{PP0+}
\end{align}
where $\mathbf{P} (0^-)$ refers to the 1PDM in the absence of the external field.

Now the crucial point is how to parameterize the local evolution operators $\mathbf{U} (t_{k + 1}, t_k)$. It should be clear from
the outset that the various low-order schemes\cite{kosloff1988time,PropagatorsKS2004,PropagatorsKS2018} are only appropriate for
the lowest part of the TDKS spectra, for they would require exceedingly small time steps to be stable and accurate enough for the high-end part
of the TDKS spectra. After extensive experimentations, it is learned here that
the oCFET4 with three exponentials\cite{oCFET2011} performs much better than MP4\cite{Magnus1954,MagnusRev} in both accuracy and efficiency.
Even higher-order propagators (MP6\cite{MagnusRev}, CFET6\cite{CFET2006}, and oCFET6\cite{oCFET2011}, etc)
are too expensive to be practical. For comparison, the closely related CFET4
with two exponentials\cite{CFET2006} is also presented here. Both oCFET4 and CFET4 are unitary by construction,
thereby conserving time reversal symmetry (in the absence of magnetic fields) as well as the trace, hermiticity, and idempotency of $\mathbf{P}(t)$.
However, strictly speaking, they are appropriate only for
cases where the Hamiltonians governing the dynamics are time independent.
For a time-dependent Hamiltonian as in RT-TDDFT, they have to be combined with some predictor-correctors
to account for the inherent non-autonomy (see Sec. \ref{SecPC}).

\subsection{CFET4 and oCFET4 Integrators}\label{SecPropagator}
The Magnus expansion\cite{Magnus1954,MagnusRev} has to be introduced first
to make the presentation of CFET\cite{CFET2006} clear. Briefly, the Magnus expansion amounts to replacing the complicated time-ordered integration
\begin{eqnarray}
\mathbf{U}(t_{k+1},t_k)=\mathcal{T} \mathrm{exp}[\int_{t_k}^{t_{k+1}} \mathbf{A}(\tau)d\tau],\quad \mathbf{A}(\tau)=-i \mathbf{F}(\tau),\label{LocalU}
\end{eqnarray}
with a time-unordered exponential of an infinite series
\begin{align}
\mathbf{U}(t_{k+1},t_k)&=\mathrm{exp}[\boldsymbol{\Omega}(t_{k+1}, t_k)]\label{PropMexp}\\
&=\mathrm{exp}[\sum_{l=1}^\infty \varepsilon^l  \boldsymbol{\Omega}_l(t_{k+1}, t_k)]_{\varepsilon=1},
\label{MagExpan2}
\end{align}
where the first two wave operators can readily be obtained as
\begin{align}
\boldsymbol{\Omega}_1(t_{k+1}, t_k) &= \int_{t_k}^{t_{k+1}}d\tau\mathbf{A}(\tau),\label{omega1 expan}\\
\boldsymbol{\Omega}_2(t_{k+1}, t_k) &= \frac{1}{2}\int_{t_k}^{t_{k+1}}d\tau_1\int_{t_k}^{\tau_1}d\tau_2[\mathbf{A}(\tau_1),\mathbf{A}(\tau_2)].\label{omega2 expan}
\end{align}
To evaluate the above time integrals, a Taylor expansion of  $\mathbf{A}(\tau)$ can first be made around the midpoint $t_{1/2}=t_k+\Delta t/2$,
\begin{equation}
\mathbf{A} (\tau) = \sum_{j = 0}^{\infty} \mathbf{a}_j (\tau - t_{1 / 2})^j,
  \quad \mathbf{a}_j = \frac{1}{j!} \left. \frac{d^j \mathbf{A}
  (\tau)}{d\tau^j} \right|_{\tau = t_{1 / 2}}. \label{taylorexpansion}
\end{equation}
Further denoting $\boldsymbol{\alpha}_i=\mathbf{a}_{i-1}\Delta t^i$, it can be shown\cite{MagnusRev} that
only $\{\boldsymbol{\alpha}_i\}_{i=1}^{s}$ are needed to construct a wave operator $\boldsymbol{\Omega}^{[2s]}$
that is correct up to order $\mathcal{O}(\Delta t^{2s})$. Specifically,
\begin{align}
  \boldsymbol{\Omega}^{[2]} (t_{k+1}, t_k) &= \boldsymbol{\alpha}_1,\label{iNME2}\\
   \boldsymbol{\Omega}^{[4]} (t_{k+1}, t_k) &= \boldsymbol{\alpha}_1 - \frac{1}{12}[\boldsymbol{\alpha}_1,\boldsymbol{\alpha}_2].\label{4th magnus}
\end{align}
To facilitate the numerical evaluation of $\{\boldsymbol{\alpha}_i\}_{i=1}^{s}$, one can first define the following moments of $\mathbf{A}(\tau)$
\begin{align}
\mathbf{A}^{(i-1)}&=\frac{1}{\Delta t^{i-1}}\int_{t_k}^{t_{k+1}} (\tau-t_{1/2})^{i-1} \mathbf{A}(\tau) d\tau,\quad i\in [1,s],\label{momenta}\\
&=\Delta t\int_0^1 (\tau-\frac{1}{2})^{i-1}\mathbf{A}(t_k+\tau\Delta t)d\tau,\\
&=\Delta t \sum_{j=1}^{n} Q_{ij}^{(s,n)}\mathbf{A}_j^{[2s]}+\mathcal{O}(\Delta t^{2s+1}),\label{mom in quad}\\
\mathbf{A}_j^{[2s]}&=\mathbf{A}(t_k+c_j^{[2s]}\Delta t),\quad Q_{ij}^{(s,n)}=w_j^{[2s]}(c_j^{[2s]}-\frac{1}{2})^{i-1},\label{Aijdef}
\end{align}
where $w_j^{[2s]}$ and $c_j^{[2s]}$ are the weights and nodes of a quadrature required to evaluate $\mathbf{A}^{(0)}$ correctly up to order $2s$, which
are applied to the other moments $\mathbf{A}^{(i)}$ as well. Inserting Eq. \eqref{taylorexpansion} into Eq. \eqref{momenta} leads to
\begin{align}
\mathbf{A}^{(i-1)}&=\sum_{j=1}^s(T^{(s)})_{ij}\boldsymbol{\alpha}_j,\nonumber\\
 (T^{(s)})_{ij}&=\frac{1+(-1)^{i+j}}{(i+j-1)2^{i+j-1}}, \quad i\in[1,s].\label{mom in alpha}
\end{align}
Combining Eqs. \eqref{mom in quad} and \eqref{mom in alpha} then gives rise to the general expression for $\{\boldsymbol{\alpha}_j\}_{j=1}^{s}$
\begin{equation}
\boldsymbol{\alpha}_j = \Delta t\sum_{k=1}^n g_{jk}\mathbf{A}_k^{[2s]},\quad g_{jk}=[(\mathbf{T}^{(s)})^{-1}\mathbf{Q}^{(s,n)}]_{jk},\quad j\in[1,s].\label{alpha in A}
\end{equation}

Consider the Gauss-Legendre quadrature. It is obvious that only one grid point is needed for
the second-order wave operator $\boldsymbol{\Omega}^{[2]}$, i.e., $w_1^{[2]}=1$ and $c_1^{[2]}=\frac{1}{2}$. We then have $Q^{(1,1)}=1$ and $(T^{(1)})^{-1}=1$ and hence
\begin{equation}
\boldsymbol{\Omega}^{[2]}(t_{k+1},t_k)=\boldsymbol{\alpha}_1=\mathbf{A}(t_k+\frac{1}{2}\Delta t)\Delta t,
\end{equation}
which gives rise to the standard, second-order midpoint Magnus propagator (MP2)
\begin{equation}
\mathbf{U}^{[2]}(t_{k+1},t_k)=\mathrm{exp}[\boldsymbol{\Omega}^{[2]}(t_{k+1},t_k)].\label{Mag2}
\end{equation}
In contrast, the evaluation of $\boldsymbol{\Omega}^{[4]}$ requires two grid points, i.e.,
 $w_1^{[4]}=\frac{1}{2}$, $c_1^{[4]}=\frac{1}{2}-\frac{\sqrt{3}}{6}$, $w_2^{[4]}=\frac{1}{2}$, and $c_2^{[4]}=\frac{1}{2}+\frac{\sqrt{3}}{6}$,
 which give rise to
\begin{equation}
\mathbf{Q}^{(2,2)} = \begin{pmatrix}
\frac{1}{2} & \frac{1}{2} \\
-\frac{\sqrt{3}}{12} & \frac{\sqrt{3}}{12}
\end{pmatrix}, \quad (\mathbf{T}^{(2)})^{-1} = \begin{pmatrix}
1 & 0\\
0 & 12
\end{pmatrix},
\end{equation}
and hence
\begin{align}
\begin{pmatrix}
\boldsymbol{\alpha}_1\\
\boldsymbol{\alpha}_2
\end{pmatrix}&=\Delta t(\mathbf{T}^{(2)})^{-1}\mathbf{Q}^{(2,2)}\begin{pmatrix}
\mathbf{A}_1^{[4]}\\
\mathbf{A}_2^{[4]}
\end{pmatrix} \nonumber\\
&= \Delta t\begin{pmatrix}
\frac{1}{2}(\mathbf{A}_1^{[4]} + \mathbf{A}_2^{[4]})\\
\sqrt{3}(\mathbf{A}_2^{[4]} - \mathbf{A}_1^{[4]})
\end{pmatrix},\\
\boldsymbol{\Omega}^{[4]}(t_{k+1},t_k) &= \frac{1}{2}(\mathbf{A}_1^{[4]} + \mathbf{A}_2^{[4]}) \Delta t - \frac{\sqrt{3}}{12}[\mathbf{A}_1^{[4]},\mathbf{A}_2^{[4]}]\Delta t^2.
\end{align}
The fourth-order Magnus propagator (MP4) then reads
\begin{align}
\mathbf{U}^{[4]}(t_{k+1},t_k)&=\mathrm{exp}[\boldsymbol{\Omega}^{[4]}(t_{k+1},t_k)]\label{Mag4}.
\end{align}

Instead of the above Magnus expansion which involves increasingly nested commutators at high orders, the commutator-free type of expansion of
Eq. \eqref{LocalU}  assumes the following form
\begin{align}
&\mathbf{U}^{[2s]}_{\mathrm{CF},m}(t_{k+1},t_k)=e^{\widetilde{\boldsymbol{\Omega}}}
= e^{\widetilde{\boldsymbol{\Omega}}_m}e^{\widetilde{\boldsymbol{\Omega}}_{m-1}}\cdots e^{\widetilde{\boldsymbol{\Omega}}_1},\\
&\widetilde{\boldsymbol{\Omega}}_i = \sum_{j = 1}^{s} f_{ij}\boldsymbol{\alpha}_j=\Delta t \sum_{k=1}^n x_{ik} \mathbf{A}_k^{[2s]},\quad x_{ik}=\sum_{j=1}^s f_{ij}g_{jk},\label{fijdef}
\end{align}
where $m$ denotes the number of exponentials.
The expansion coefficients $f_{ij}$ can be determined\cite{CFET2006} by equating
the terms of $\widetilde{\boldsymbol{\Omega}}$  to those of the Magnus expansion $\boldsymbol{\Omega}$ \eqref{PropMexp} at each order, up to $\mathcal{O}(\Delta t^{2s})$.
Consider CFET4 with two exponentials
\begin{equation}
\mathbf{U}_{\mathrm{CF},2}^{[4]}(t_{k+1},t_{k}) = e^{f_{21}\boldsymbol{\alpha}_1+f_{22}\boldsymbol{\alpha}_2}e^{f_{11}\boldsymbol{\alpha}_1+f_{12}\boldsymbol{\alpha}_2},
\end{equation}
where the coefficients should be subject to
\begin{equation}
f_{21}=f_{11},\quad f_{22}=-f_{12}\label{trs of cfet}
\end{equation}
to fulfill time reversal symmetry. Further use of
the Baker-Campbell-Hausdorff formula can be made to merge the two exponentials into one
\begin{equation}
\mathbf{U}_{\mathrm{CF},2}^{[4]}(t_{k+1},t_k) = e^{2f_{11}\boldsymbol{\alpha}_1+\frac{f_{12}}{2}[\boldsymbol{\alpha}_1,\boldsymbol{\alpha}_2]},
\end{equation}
the exponent of which can be equated to $\boldsymbol{\Omega}^{[4]} (t_{k+1}, t_k)$ in Eq. \eqref{4th magnus}, thereby leading to
\begin{equation}
\mathbf{U}_{\mathrm{CF},2}^{[4]}(t_{k+1},t_k) = e^{\frac{1}{2}\boldsymbol{\alpha}_1+\frac{1}{6}\boldsymbol{\alpha}_2}e^{\frac{1}{2}\boldsymbol{\alpha}_1-\frac{1}{6}\boldsymbol{\alpha}_2},
\end{equation}
the proper implementation of which is described in Algorithm \ref{mag42} in Appendix \ref{AppA}.

Instead of the termwise matching between CFETn and MPn, the expansion coefficients $\{f_{ij}\}$ in Eq. \eqref{fijdef}
can also be optimized\cite{oCFET2011} to minimize the error between the approximate $\widetilde{\boldsymbol{\Omega}}$ and the exact $\boldsymbol{\Omega}$, thereby leading to the optimized (oCFET) variants of CFET. Without going into further details, the required coefficients $\{x_{ik}\}$ can be determined by reexpressing
the coefficients $\mathbf{A}_n$ defined in equation (25) of Ref. \citenum{oCFET2011} in terms of the present moments $\mathbf{A}_j^{[2s]}$ \eqref{Aijdef}.
The oCFET4 with three exponentials considered here is specified in Algorithm \ref{mag43} in Appendix \ref{AppA}.

\subsection{Predictor-Correctors}\label{SecPC}
The above CFET4 and oCFET4 have to be combined with suitable predictor-correctors to account for
the time dependence in RT-TDDFT. The available predictor-correctors \cite{PropagatorsKS2004,Voorhis06,Herbert2018,RT4CTD2015}
were designed only for MP2 and hence cannot be used here. Instead, we got to design new predictor-correctors for CFET4 and oCFET4 case by case.

\subsubsection{Predictor-Corrector for CFET4}

As a fourth-order propagator, CFET4 requires two Gauss-Legendre quadrature points. For this reason,
an EPEP2 (exponential prediction of density matrix and exponential correction of density matrix at two Gauss-Legendre points) type of predictor-corrector
is proposed here, see Fig. \ref{EPEP2} (and Algorithm \ref{prop4 implicit exp} in Appendix \ref{AppA}). The workflow goes as follows.

\begin{inparaenum}[\quad Step (1):]
\item Propagate $\mathbf{P}(t_k)$ to $\mathbf{P}(t_1^{[4]})$ with the second-order enforced time reversal symmetry (ETRS) propagator\cite{PropagatorsKS2004} (see Algorithm \ref{etrs1} in Appendix \ref{AppA}),
\begin{align}
  \mathbf{U}_{\mathrm{ETRS}} (t_{k+1}, t_k) & = e^{- i \frac{\Delta t}{2} \mathbf{F} (t_{k+1})} e^{- i \frac{\Delta t}{2} \mathbf{F} (t_k)},\label{etrsprop}\\
  \mathbf{P}(t_1^{[4]}) & = \mathbf{U}_{\mathrm{ETRS}}(t_{k+1},t_k)\mathbf{P}(t_k)\mathbf{U}_{\mathrm{ETRS}}^{\dagger}(t_{k+1},t_k), \nonumber\\
  t_1^{[4]} &=t_k + c_1^{[4]}\Delta t.
\end{align}

\item Construct $\mathbf{F}(t_1^{[4]})$ with $\mathbf{P}(t_1^{[4]})$.

\item Propagate $\mathbf{P}(t_1^{[4]})$ to $\mathbf{P}(t_2^{[4]})$ with Euler's rule,
\begin{align}
\mathbf{P}(t_2^{[4]}) &= e^{-i\mathbf{F}(t_1^{[4]})(t_2^{[4]}-t_1^{[4]})}\mathbf{P}(t_1^{[4]})e^{i\mathbf{F}(t_1^{[4]})(t_2^{[4]}-t_1^{[4]})},\nonumber \\
 t_2^{[4]}&=t_k + c_2^{[4]}\Delta t.\label{t42-t41Euler}
\end{align}

\item\label{Iter40} Construct $\mathbf{F}(t_2^{[4]})$ with $\mathbf{P}(t_2^{[4]})$.

\item Propagate $\mathbf{P}(t_k)$ to $\mathbf{P}(t_{k+1})$ with CFET4 (see Algorithm \ref{mag42} in Appendix \ref{AppA}).

\item Construct $\mathbf{F}(t_{k+1})$ with $\mathbf{P}(t_{k+1})$.

\item\label{Iter41} Backward propagate $\mathbf{P}(t_{k+1})$ to $\mathbf{P}(t_2^{[4]})$ with Euler's rule,
\begin{equation}
\mathbf{P}(t_2^{[4]}) = e^{i\mathbf{F}(t_{k+1})(t_{k+1}-t_2^{[4]})}\mathbf{P}(t_{k+1})e^{-i\mathbf{F}(t_{k+1})(t_{k+1}-t_2^{[4]})},
\end{equation}
so as to correct $\mathbf{P}(t_2^{[4]})$.

\item Repeat steps (\ref{Iter40}) to (\ref{Iter41}) until $\mathbf{P}(t_{k+1})$ converges according to the following criterion\cite{Herbert2018}
\begin{equation}
\frac{\Vert\mathbf{X}_1-\mathbf{X}_2\Vert_F}{M}<\xi,\label{converge criterion}
\end{equation}
where $\Vert \cdot \Vert_F$ denotes the Frobenius norm, $\mathbf{X}_1$ and $\mathbf{X}_2$ represents two $M\times M$ matrices to be compared, and $\xi$ is the chosen threshold.
\end{inparaenum}

In the above procedure, $\mathbf{P}(t_1^{[4]})$ does not enter the iteration cycle (steps (\ref{Iter40}) to (\ref{Iter41})) but
is predicted directly from $\mathbf{P}(t_k)$. That is, $\mathbf{P}(t_1^{[4]})$ is not to be corrected iteratively by $\mathbf{P}(t_{k+1})$ although this can in principle be done.
Such an option arises from the following consideration: $t_1^{[4]}$ is close to $t_k$, such that
the second-order ETRS (instead of the first-order Euler) propagator would ensure an accurate prediction of $\mathbf{P}(t_1^{[4]})$. In contrast,
had $\mathbf{P}(t_1^{[4]})$ been corrected by $\mathbf{P}(t_{k+1})$, the accuracy would be deteriorated due to the larger interval between $t_1^{[4]}$ and $t_{k+1}$.

\begin{figure}[h]
\centering
\includegraphics[width=0.5\textwidth]{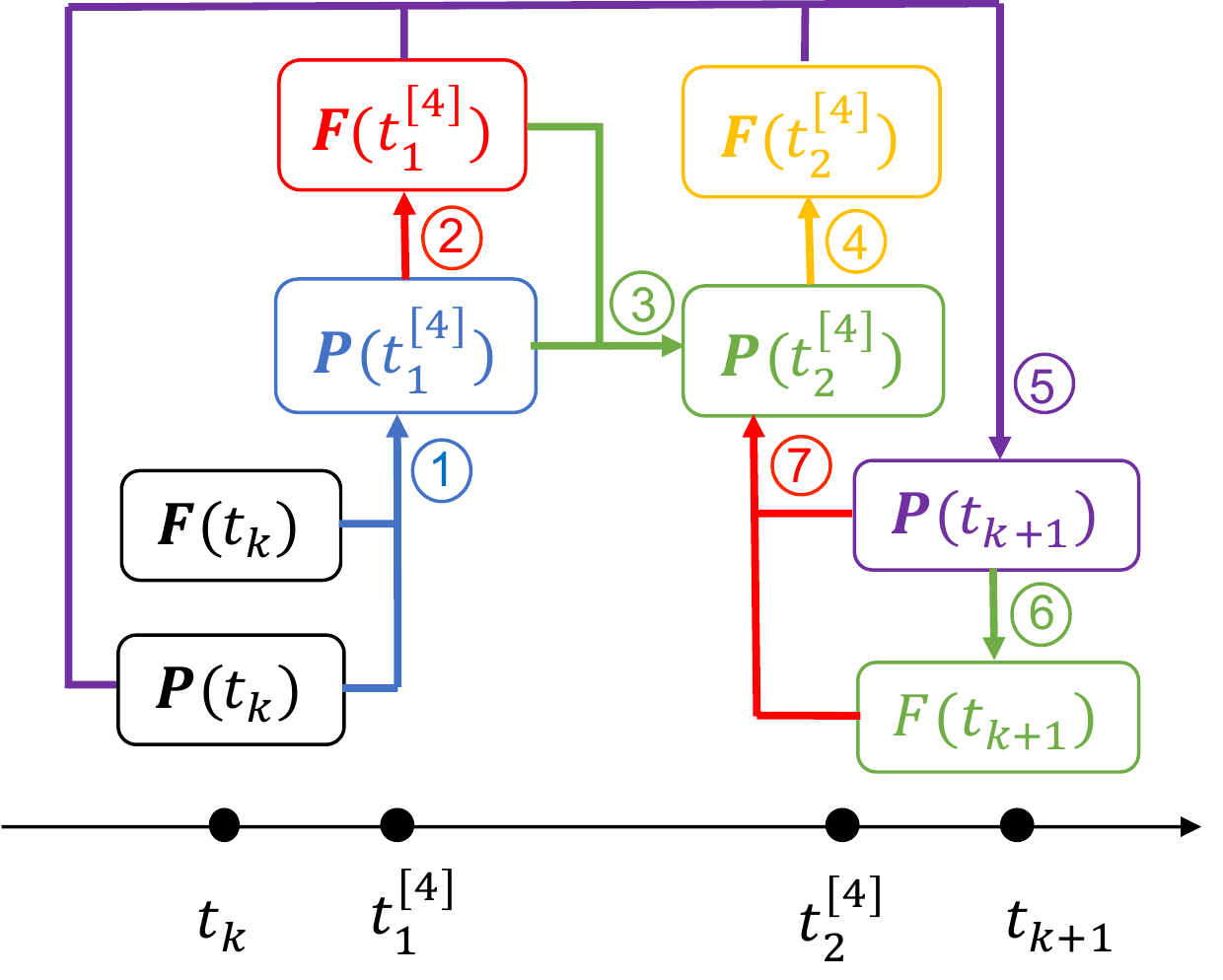}
\caption{One time-step flowchart of EPEP2 for CFET4. In circles are step indices of Algorithm \ref{prop4 implicit exp} in Appendix \ref{AppA}.}\label{EPEP2}
\end{figure}

\subsubsection{Predictor-Corrector for oCFET4}

Albeit a fourth-order propagator, oCFET4 requires three Gauss-Legendre quadrature points. In the same spirit as EPEP2, an EPEP3 type of predictor-corrector
is proposed here, see Fig. \ref{EPEP3} (and Algorithm \ref{prop6 implicit exp} in Appendix \ref{AppA}). The workflow goes as follows.

\begin{inparaenum}[\quad Step 1:]
\item Propagate $\mathbf{P}(t_k)$ to $\mathbf{P}(t_1^{[6]})$ with ETRS. Here, $t_1^{[6]}= t_k+c_1^{[6]}\Delta t$ and $c_1^{[6]}=\frac{1}{2}-\frac{\sqrt{15}}{10}$.

\item Construct $\mathbf{F}(t_1^{[6]})$ with $\mathbf{P}(t_1^{[6]})$.

\item Propagate $\mathbf{P}(t_1^{[6]})$ to $\mathbf{P}(t_2^{[6]})$ with Euler's rule,
\begin{align}
\mathbf{P}(t_2^{[6]}) &= e^{-i\mathbf{F}(t_1^{[6]})(t_2^{[6]}-t_1^{[6]})}\mathbf{P}(t_1^{[6]})e^{i\mathbf{F}(t_1^{[6]})(t_2^{[6]}-t_1^{[6]})}, \nonumber\\
 t_2^{[6]}&= t_k+c_2^{[6]}\Delta t,\quad c_2^{[6]}=\frac{1}{2}.\label{t62-t61Euler}
\end{align}

\item Construct $\mathbf{F}(t_2^{[6]})$ with $\mathbf{P}(t_2^{[6]})$.

\item Propagate $\mathbf{P}(t_1^{[6]})$ to $\mathbf{P}(t_3^{[6]})$ with MP2 \eqref{Mag2},
\begin{align}
\mathbf{P}(t_3^{[6]}) &= e^{-i\mathbf{F}(t_2^{[6]})(t_3^{[6]}-t_1^{[6]})}\mathbf{P}(t_3^{[6]})e^{i\mathbf{F}(t_2^{[6]})(t_3^{[6]}-t_1^{[6]})}, \nonumber\\
 t_3^{[6]}&= t_k+c_3^{[6]}\Delta t, \quad c_3^{[6]}=\frac{1}{2}+\frac{\sqrt{15}}{10}.\label{t63-t61M2}
\end{align}

\item Construct $\mathbf{F}(t_3^{[6]})$ with $\mathbf{P}(t_3^{[6]})$.

\item\label{Start} Propagate $\mathbf{P}(t_k)$ to $\mathbf{P}(t_{k+1})$ with oCFET4 (see Algorithm \ref{mag43} in Appendix \ref{AppA}).

\item Construct $\mathbf{F}(t_{k+1})$ with $\mathbf{P}(t_{k+1})$.

\item Backward propagate $\mathbf{P}(t_{k+1})$ to $\mathbf{P}(t_3^{[6]})$ with Euler's rule,
\begin{equation}
\mathbf{P}(t_3^{[6]})  = e^{i\mathbf{F}(t_{k+1})(t_{k+1}-t_3^{[6]})}\mathbf{P}(t_{k+1})e^{-i\mathbf{F}(t_{k+1})(t_{k+1}-t_3^{[6]})},
\end{equation}
so as to correct $\mathbf{P}(t_3^{[6]})$.

\item Construct $\mathbf{F}(t_3^{[6]})$ with $\mathbf{P}(t_3^{[6]})$.

\item Backward propagate $\mathbf{P}(t_3^{[6]})$ to $\mathbf{P}(t_4^{[6]})$ with Euler's rule,
\begin{align}
\mathbf{P}(t_4^{[6]}) &= e^{i\mathbf{F}(t_3^{[6]})(t_3^{[6]}-t_4^{[6]})}\mathbf{P}(t_3^{[6]})e^{-i\mathbf{F}(t_3^{[6]})(t_3^{[6]}-t_4^{[6]})},\nonumber\\
t_4^{[6]}&= \frac{t_2^{[6]}+t_3^{[6]}}{2}.
\end{align}

\item Construct $\mathbf{F}(t_4^{[6]})$ with $\mathbf{P}(t_4^{[6]})$.

\item Backward propagate $\mathbf{P}(t_3^{[6]})$ to $\mathbf{P}(t_2^{[6]})$ with MP2 \eqref{Mag2},
\begin{equation}
\mathbf{P}(t_2^{[6]}) = e^{i\mathbf{F}(t_4)(t_3^{[6]}-t_2^{[6]})}\mathbf{P}(t_3^{[6]})e^{-i\mathbf{F}(t_4)(t_3^{[6]}-t_2^{[6]})},
\end{equation}
so as to correct $\mathbf{P}(t_2^{[6]})$.

\item\label{End} Construct $\mathbf{F}(t_2^{[6]})$ with $\mathbf{P}(t_2^{[6]})$.

\item Repeat steps (\ref{Start}) to (\ref{End}) until $\mathbf{P}(t_{k+1})$ converges according to criterion \eqref{converge criterion}.
\end{inparaenum}

\begin{figure}[h]
\centering
\includegraphics[width=0.5\textwidth]{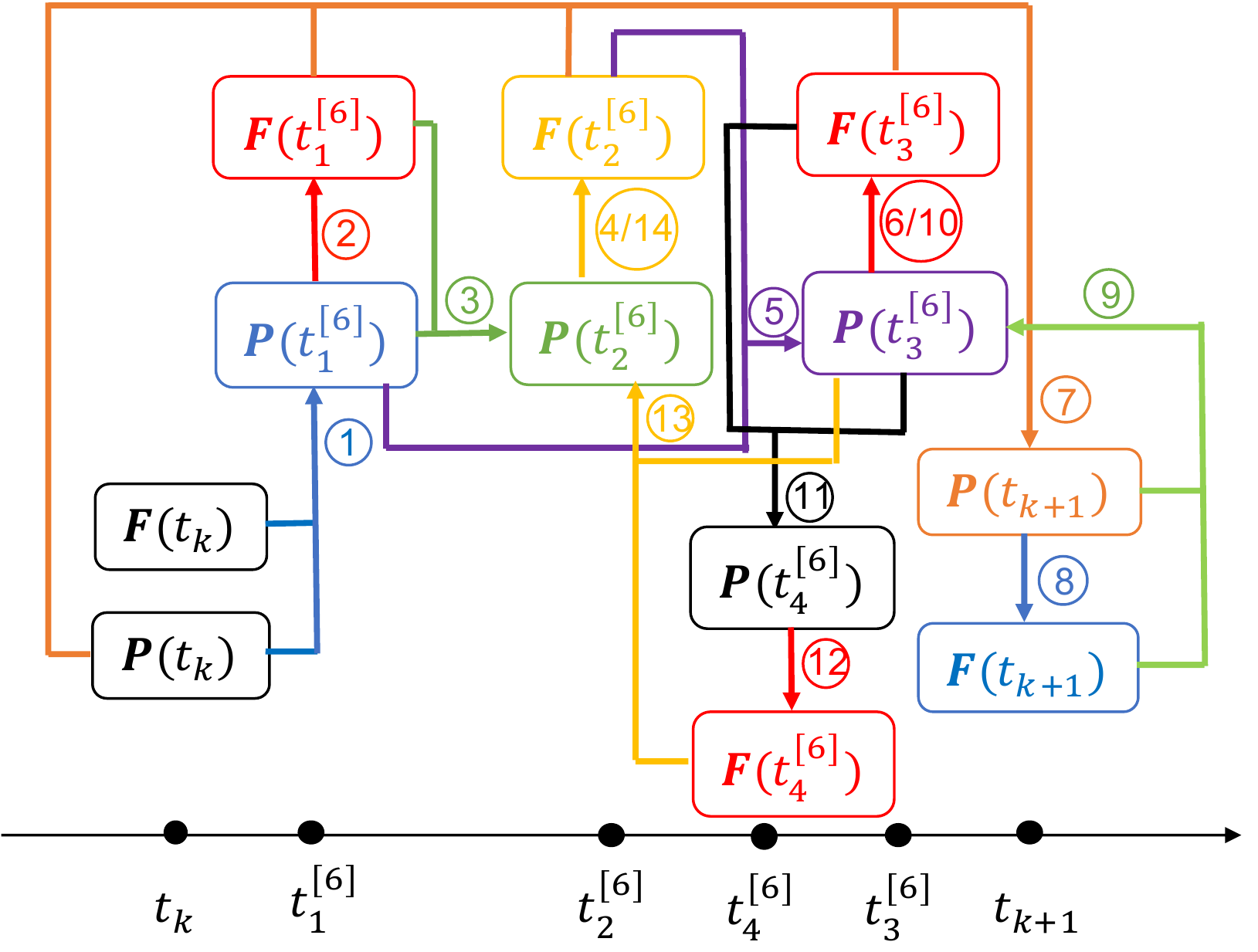}
\caption{One time-step flowchart of EPEP3 for oCFET4. In circles are step indices of Algorithm \ref{prop6 implicit exp} in Appendix \ref{AppA}.}\label{EPEP3}
\end{figure}

In the above procedure, $\mathbf{P}(t_1^{[6]})$ is predicted directly from $\mathbf{P}(t_k)$, for $t_1^{[6]}$ is much closer to $t_k$ than
to $t_{k+1}$. This stays in the same spirit as EPEP2. $\mathbf{P}(t_3^{[6]})$ is corrected by
using the backward Euler propagation of $\mathbf{P}(t_{k+1})$, for $t_3^{[6]}$ is closer to $t_{k+1}$.
Since the interval $t_3^{[6]}-t_2^{[6]}$ for $\mathbf{P}(t_2^{[6]})$  is
larger than the interval $t_{k+1}-t_3^{[6]}$ for $\mathbf{P}(t_3^{[6]})$,
an auxiliary time moment $t_4^{[6]}$ has been introduced to correct $\mathbf{P}(t_2^{[6]})$ by using
the backward MP2 propagation from $\mathbf{P}(t_3^{[6]})$.

Since EPEP3 and EPEP2 share the same strategy, they should converge at the same rate.
As a matter of fact, in the present calculations of core excitations with the predicted time steps (see Sec. \ref{SecStep}),
the first iteration already gives rise to accurate predictions of $\mathbf{P}(t_{k+1})$, such that the second iteration is just
to confirm the convergence, which is far below the chosen threshold $\xi=10^{-7}$ in Eq. \eqref{converge criterion}.

\subsection{Time Step}\label{SecStep}
\begin{figure}[h]
\includegraphics[width=0.5\textwidth]{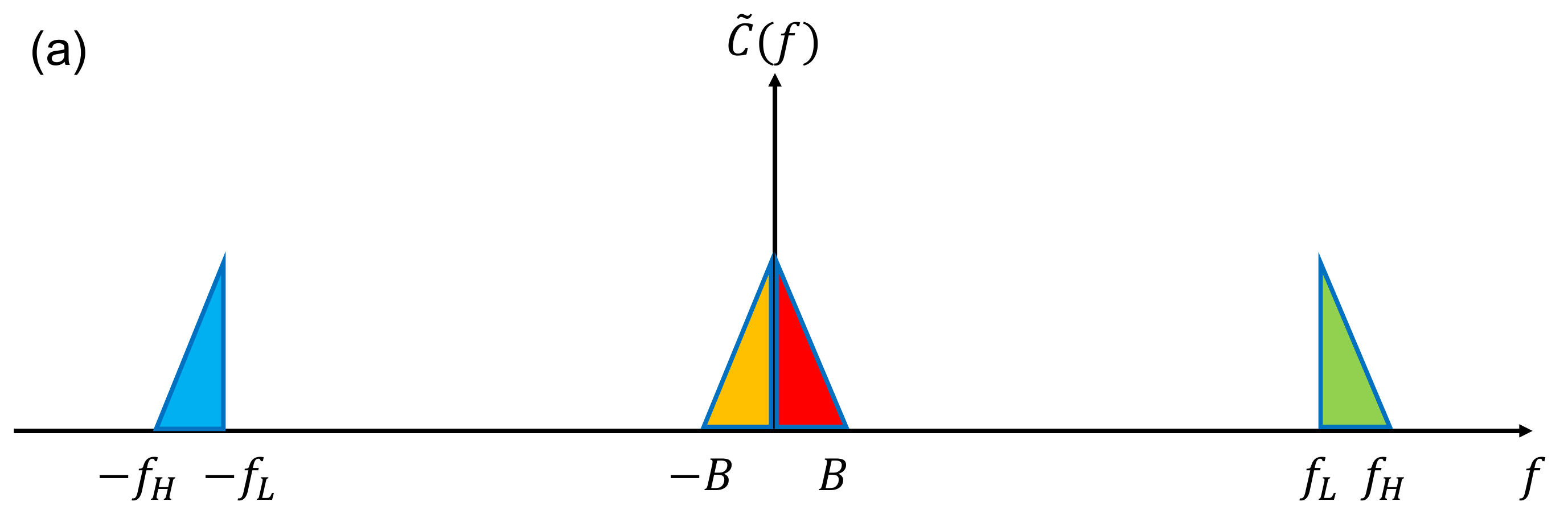}
\includegraphics[width=0.5\textwidth]{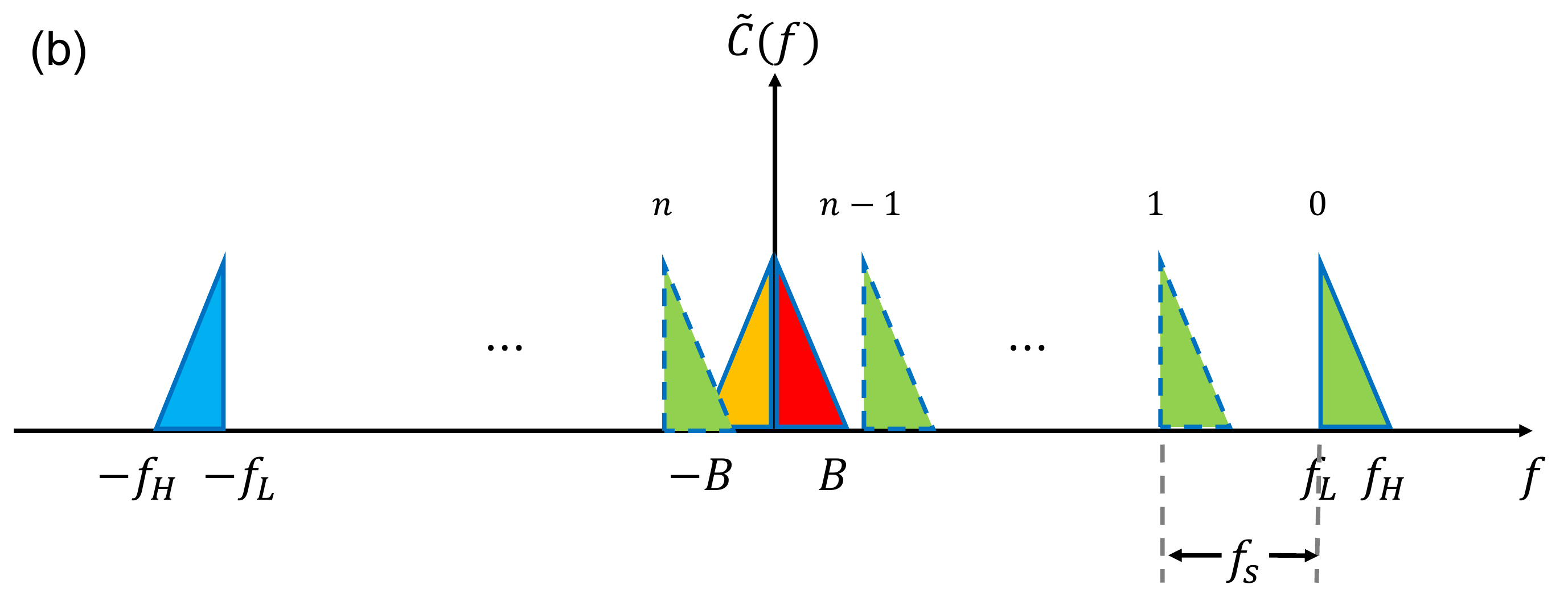}
  \caption{\label{alias} (a) Schematic representation of a sparse spectrum, with frequencies located in blocks $[- f_H, -
  f_L]$, $[- B, 0]$, $[0, B]$, and $[f_L, f_H]$.
  (b) Green blocks with dashed contour lines are periodic images (aliases) of the original
  frequencies in $[f_L, f_H]$, separated by the sampling frequency
  $f_s$. The conditions in Eq. \eqref{noaliasinB5} ensure that the $n$-th and $(n-1)$-th image blocks of $[f_L, f_H]$
  are located on the left and right hands of sides of $[0,B]$, respectively. Further incorporation of the conditions in Eq. \eqref{noaliasinB7}
  will separate the $n$-th image block of $[f_L, f_H]$ from $[-B,0]$, as a result of the separation between $[0,B]$ and the $n$-th image block of $[-f_H,-f_L]$
  (because $[-f_H,-f_L]$ and $[f_L, f_H]$ are symmetric with respect to the 0 Hz axis). }
\end{figure}

Having chosen an appropriate propagator (e.g., MP2 for low-lying valence excitations and oCFET4 for core excitations),
along with the corresponding corrector-predictor, the next question is how to choose an appropriate time step $\Delta t$.
Instead of playing around with $\Delta t$ in a trial-and-error manner, we here try to fix it in an \emph{a priori} manner, so as to facilitate the dynamics simulations.
It is well known that $\Delta t$ is bounded by both the stability and accuracy of the chosen propagator as well as the Nyquist criterion
for signal sampling\cite{Nyquist1928,shannon1948}, i.e.,
$\Delta t \le \Delta t_H = \frac{1}{f_s}< \frac{1}{2 f_H}$, with $f_H$ being the highest temporal frequency contained in the sampled signal.
Literally, the sampling frequency $f_s$ ($=1/\Delta t$) must be twice larger than the width $f_H-(-f_H)$ of the spectrum contained in the band limited real signal.
This restriction is to avoid the so-called aliasing phenomenon, which stems from
the fact that every frequency within $[-f_H, f_H]$ will be copied infinitely many times (separated by the sampling frequency $f_s$)
by the Fourier transform of the sampled signal.
However, if the spectrum $[-f_H, f_H]$ is separated in blocks and only one particular block is of interest,
an even larger time step $\Delta t$ can be chosen to sample the signal. To see this,
consider the spectrum of a real-valued time signal shown in Fig. \ref{alias}, which has
four blocks, $[- f_H, - f_L]$, $[- B, 0]$, $[0, B]$, and $[f_L, f_H]$. To extract the frequencies within $[0, B]$,
the proper sampling frequency $f_s$
can be determined by the requirement that no alias from the other blocks will appear in the interval $[0, B]$. Consider first
the block $[f_L,f_H]$, which is shifted to the left by $(n-1)$ times until the right end of $[0, B]$. The next left shift
of $[f_L,f_H]$ must cross $[0, B]$ in one step of $f_s$, thereby resulting in the first necessary condition for $f_s$,
\begin{equation}
  f_s > B + f_H - f_L.\label{noaliasinB}
\end{equation}
The same is obtained by right shifts of $[-f_H,-f_L]$ in steps of $f_s$. Likewise, the one-step right shift of $[-B,0]$ gives rise to the following necessary condition
\begin{equation}
f_s > 2B.\label{noaliasinB2}
\end{equation}
The two conditions \eqref{noaliasinB} and \eqref{noaliasinB2} can be combined to yield the lower bound $\tilde{f}_s$ of $f_s$
\begin{equation}
\tilde{f}_s=\max\{B+f_H-f_L, 2B\}<f_s.\label{noaliasinB3}
\end{equation}
Moreover, it is obvious that the following conditions
\begin{eqnarray}
  f_L - (n - 1) f_s  >  B, \quad
  f_H - \mathrm{nf}_s  <  0, \quad n=1,2,\cdots,\label{Flhcondb}
\end{eqnarray}
should also be satisfied simultaneously for the left shifts of $[f_L,f_H]$. The conditions in Eqs. \eqref{noaliasinB3} and \eqref{Flhcondb} can be
combined to yield
\begin{eqnarray}
  \frac{f_H }{n}  < f_s <  \frac{f_L - B}{n - 1},  \quad
  1  \leqslant n \leqslant \left\lfloor \frac{f_L - B}{\tilde{f}_s}
  \right\rfloor + 1,  \label{noaliasinB5}
\end{eqnarray}
where the symbol $\lfloor y \rfloor$ denotes the largest integer less than or equal to $y$.
Similarly, the following conditions
\begin{eqnarray}
  -f_L + (m - 1) f_s  <  0, \quad
  -f_H + mf_s  >  B, \quad m=1,2,\cdots,\label{Flhcondd}
\end{eqnarray}
should be satisfied additionally for the right shifts of $[-f_H,-f_L]$. The conditions in Eqs. \eqref{noaliasinB3} and \eqref{Flhcondd} can be
combined to yield

\begin{eqnarray}
  \frac{f_H+B}{m}  < f_s <  \frac{f_L}{m - 1},\quad
  1  \leqslant m \leqslant \left\lfloor \frac{f_L}{\tilde{f}_s} \right\rfloor + 1.  \label{noaliasinB7}
\end{eqnarray}
Finally, no additional condition is required for the right shift of $[-B,0]$.

In summary, to read out correctly the frequencies in the interval $[0,B]$ from the Fourier transform of the sampled signal,
the sampling frequency $f_s$ has to satisfy the block aliasing conditions (BAC) \eqref{noaliasinB3}, \eqref{noaliasinB5}, and \eqref{noaliasinB7} simultaneously.
The analysis can readily be generalized to extract the frequencies of any block of a spectrum composed of $2N$ blocks.
The algorithm goes as follows.
\begin{enumerate}[(1)]
\item For each symmetry $\Gamma$ of the excited states, determine the corresponding symmetries of the virtual ($\gamma_p$) and occupied ($\gamma_h$) canonical molecular orbitals (CMOs) and calculate their energy differences $\{\Delta \epsilon_{ai}\}|_{a\in \gamma_p, i\in \gamma_h}$.
Sort $\{\Delta \epsilon_{ai}\}$ of all the $\gamma_p$-$\gamma_h$ pairs
in ascending order, leading to $\{\Delta\epsilon_{n}\}_{n=1}^{\Gamma,N_{\Gamma}}$.

\item  Partition $\{\Delta\epsilon_{n}\}_{n=1}^{\Gamma,N_{\Gamma}}$ into blocks $\lceil\Delta\epsilon\rceil^{\Gamma,M_\Gamma}_{m=1}$
according to their gaps. Specifically, if the separation between $\Delta\epsilon_{n}$ and $\Delta\epsilon_{n+1}$ is larger than $\omega_{G}=2\omega_b+\omega_{c}$,
set $\Delta\epsilon_n$ and $\Delta\epsilon_{n+1}$ as the uppermost and lowermost elements of the current and next blocks, respectively.
Here, $\omega_b$ is a parameter (e.g., 0.5 a.u.) to broaden each block, so as to account for the uncertainties of the independent particle approximation (IPA) of the true excitation energies,
whereas $\omega_c$ is another parameter (e.g., 0.2 a.u.) to ensure the minimal separation between two adjacent broadened blocks (see Fig. S1/S2 for an example). The in total $M_{\Gamma}$ blocks $\frac{1}{2\pi}\lceil\Delta\epsilon\rceil^{\Gamma,M_{\Gamma}}_{m=1}\equiv[f_{Lm},f_{Hm}]_{m=1}^{\Gamma,M_{\Gamma}}$
along with their negative counter parts $[-f_{Hm},-f_{Lm}]_{m=1}^{\Gamma,M_{\Gamma}}$ form the estimated structure of the whole TDKS spectrum .

\item  Choose a target block, say $[f_{Lk},f_{Hk}]$, and calculate the minimal sampling frequency
\begin{equation}
\tilde{f}_s = f_{Hk}-f_{Lk}+\max_{m\in[1,M_{\Gamma}]}(f_{Hm}-f_{Lm}).\label{fsmin}
\end{equation}

\item Build up the corresponding inequalities for each block $[f_a,f_b]$ ($f_b>f_a$; including those of negative frequencies). For those blocks lower and higher than $[f_{Lk},f_{Hk}]$, we have
\begin{eqnarray}
\frac{f_{Hk}-f_a}{n} < f_s < \frac{f_{Lk}-f_b}{n-1}, \quad
1  \leqslant n \leqslant  \left\lfloor \frac{f_{Lk}-f_b}{\tilde{f}_s} \right\rfloor + 1,
\end{eqnarray}
and
\begin{eqnarray}
\frac{f_b-f_{Lk}}{n} <f_s< \frac{f_a-f_{Hk}}{n-1},\quad
1  \leqslant n \leqslant  \left\lfloor \frac{f_a-f_{Hk}}{\tilde{f}_s} \right\rfloor + 1,
\end{eqnarray}
respectively.
\item  Determine the intersection of the inequalities of all blocks so as to obtain the allowed sampling frequency $f_s$ and hence the time step $\Delta t=1/f_s$ for each symmetry $\Gamma$ of excited states.
\end{enumerate}
Note that when the whole spectrum is regarded as one big block $[-f_H,f_H]$, the above algorithm reduces naturally to the Nyquist time step $\Delta t_H=\frac{1}{f_s}=\frac{1}{2f_H}$, as can be seen from Eq. \eqref{fsmin}.

\begin{figure}[h]
\centering
\begin{tabular}{cc}
\includegraphics[height=7cm]{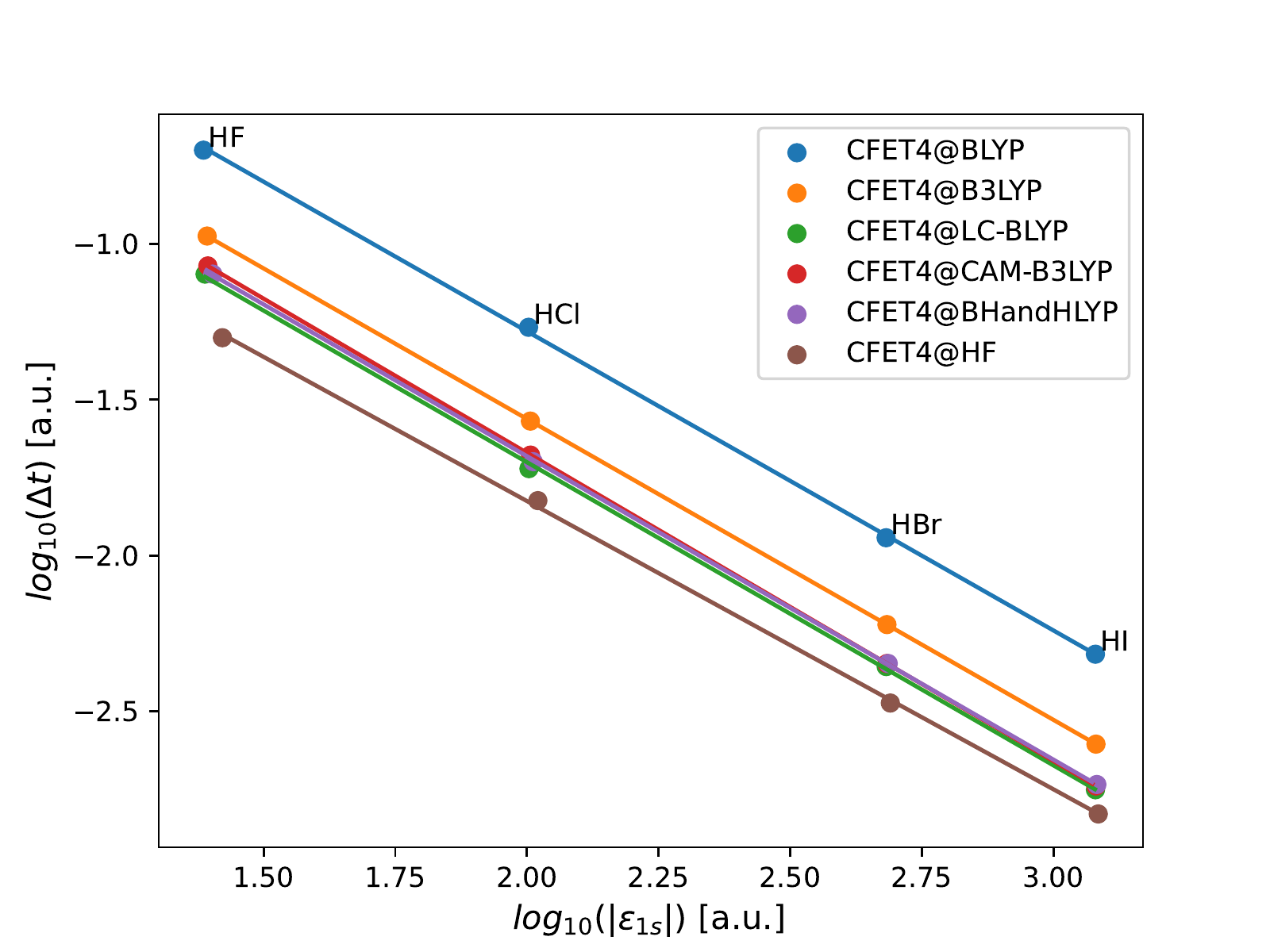}\\
(a)\\
\includegraphics[height=7cm]{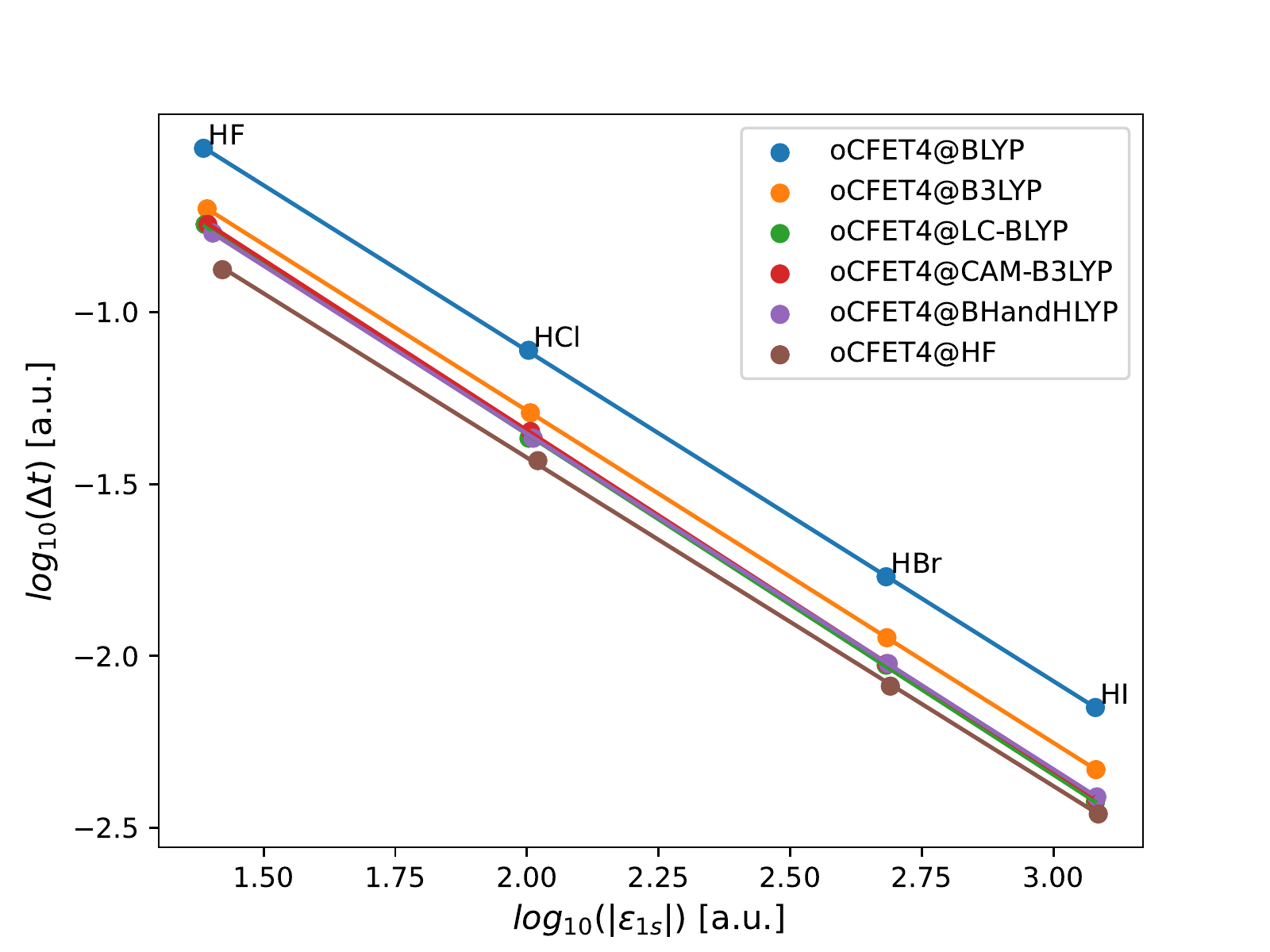}\\
(b)
\end{tabular}
\caption{Linear relations between $log_{10}(\Delta t)$ and $log_{10}(|\epsilon_{1s}|)$ for the lowest five bright $K$-edge core excited states of HF, HCl, and HBr by (a) CFET4@RT-TDDFT and (b) oCFET4@RT-TDDFT
with different exchange-correlation functionals (nonrelativistic calculations with the DEF2-SVPD basis set\cite{DEF2SVPD05,DEF2SVPD10}).
$\Delta t$: the largest time step for an accuracy of $<0.05$ eV; $\epsilon_{1s}$: the ground-state lowest core orbital energy.}\label{deltat_ene fit}
\end{figure}

At this stage, it must be realized that the time step determined by the above algorithm only
guarantees that the target frequencies are not messed up with aliases from the other blocks.
The accuracy of the spectral analysis depends further on the quality of the input time signal.
For a given propagator, the smaller the time step, the more accurate the time signal. On the other hand, for a given time step,
the more accurate the propagator, the more accurate the time signal. However, a very small time step or a very accurate propagator usually
implies a very high computational cost. Therefore, it is important to balance these factors
in the allowed range of sampling frequency. To achieve this,  we first tried using the following parameter
\begin{equation}
\tau_{P} = \frac{\Vert\mathbf{P}(t)-\mathbf{P}_{\mathrm{ref}}(t)\Vert_{F}}{M\cdot t}
\end{equation}
to quantify the accuracy of $\mathbf{P}(t)$ (evolved by the chosen propagator with time step $\Delta t$) against a reference 1PDM $\mathbf{P}_{\mathrm{ref}}(t)$
(obtained by CF6:5Opt (see Table 6 in Ref.\citenum{oCFET2011}) with $\Delta t_H/2$).
However, preliminary experimentation reveals that, for core excitations, $\tau_{P}$ is rather sensitive to all the ingredients
(atomic number, XC functional, electron kinematics, and basis set), meaning that it is not directly connected
to the target accuracy. Having realized that such ingredients have been `renormalized' into the lowest core level $\epsilon_{1s}$ of the ground-state calculation,
we switch to seek for a relation between $\Delta t$ and $\epsilon_{1s}$ for core excitations. As can be seen from Fig. \ref{deltat_ene fit},
for the lowest five bright $K$-edge core excited states of HX (X = F, Cl, Br),
there exist very good linear relations between $log_{10}(\Delta t)$ and $log_{10}(|\epsilon_{1s}|)$ for both CFET4 and oCFET4.
Here, $\Delta t$  is the largest time step to guarantee the desired spectral accuracy of less than 0.05 eV. The fitting functions for different XC functionals
are further documented in Table \ref{LinFit}. Noticeably, the slopes and interceptions are slightly different for different XC functionals
but which are hardly surprising, for different functionals
leads to different degrees of noncommutativity between KS matrices at different times: the more exact exchange, the more enhanced noncommutativity
and hence the smaller the time step.
Although such relations have been obtained based on nonrelativistic calculations with a single basis set (DEF2-SVPD\cite{DEF2SVPD05,DEF2SVPD10}), they should be applicable
to other basis sets and other systems with relativistic effects accounted for. The underlying reasons are twofold: (1) different basis sets
produce essentially the same energies for all the occupied and low-lying virtual orbitals and (2) relativity is heavily dominated by
the time-independent one-body terms (which do not cause noncommutativity between KS matrices at different times).
In other words, such linear relations are very robust with respect to atomic number,
electron kinematics, and basis set, and can hence be applied to all the elements in the Periodic Table,
to achieve an accuracy better than 0.05 eV for core excitation energies as compared to those by LR-TDDFT.
To confirm this, the time steps predicted by the linear relations in Table \ref{LinFit} (which requires only the ground-state lowest core orbital energy
$\epsilon_{1s}$ as input)
will be taken in Sec. \ref{Results} for the $K$- and $L$-edge core excitations of HI and PbO.
Note in passing that the time step predicted by the linear relations should further be verified to comply with the above BACs.
If not, it should be reduced to the largest allowed value.

After having determined the time step for each propagator, it is attempting to know which propagator
will be most efficient in terms of the number of Fock builds. To this end,
the numbers ($N_{\mathrm{FBS}}$) of Fock builds per time step are summarized in Table \ref{Fockbuild} for MP2, MP4, CFET4, and oCFET4.

\begin{table}[h]
\centering
\caption{Fitting functions$^*$ $log_{10}(\Delta t)=a \cdot log_{10}(|\epsilon_{1s}|)+b$ for $K$-edge core excitations by CFET4@RT-TDDFT and oCFET4@RT-TDDFT (cf.
 Fig. \ref{deltat_ene fit}) }
 \small
\begin{threeparttable}
\begin{tabular}{cllllll}\toprule
\multirow{1}{*}{propagator}& functional &\multicolumn{1}{c}{$a$} &\multicolumn{1}{c}{$b$} \\
\toprule
\multirow{6}{*}{CFET4}
& BLYP\cite{B88x,LYP} & -0.96018601 & 0.63991532 \\
& B3LYP\cite{Becke93,B3LYP} & -0.96592829 & 0.37019600 \\
& LC-BLYP\cite{LCBLYP} & -0.97307440 & 0.24557525 \\
& CAM-B3LYP\cite{CAMB3LYP} & -0.98942182 & 0.30841338 \\
& BHandHLYP\cite{Becke93,B88x,LYP} & -0.97407896 & 0.26693596 \\
& HF  &  -0.92526295 & 0.02526600\\ \\
\multirow{6}{*}{oCFET4}
& BLYP\cite{B88x,LYP} & -0.96171191 & 0.81185106 \\
& B3LYP\cite{Becke93,B3LYP} & -0.96656731 & 0.64711096 \\
& LC-BLYP\cite{LCBLYP} & -0.99080653 & 0.62700701 \\
& CAM-B3LYP\cite{CAMB3LYP} & -0.99061081 & 0.63794424 \\
& BHandHLYP\cite{Becke93,B88x,LYP} & -0.97647005 & 0.59990197 \\
& HF  &  -0.95558724 & 0.48821463\\
\bottomrule
\end{tabular}
\begin{tablenotes}
\item[*] $\epsilon_{1s}$: the ground-state lowest core orbital energy.
\end{tablenotes}
\end{threeparttable}
\label{LinFit}
\end{table}

\begin{table}[h]
\centering
\caption{Numbers ($N_{\mathrm{FBS}}$) of Fock builds per time step involved in the predictor-correctors}
\small
\begin{threeparttable}
\begin{tabular}{ccccccc}
   \toprule
    Predictor-Corrector & Propagator & $N_{\mathrm{FBS}}$\tnote{a} & Formal order \\
    \midrule
     EPEP2\tnote{b} & CFET4, MP4 & $3 \cdot N_{\mathrm{it}} $ & 4 \\
     EPEP3\tnote{c} & oCFET4 & $5 \cdot N_{\mathrm{it}}-1$ & 4 \\
    \bottomrule
\end{tabular}
   \begin{tablenotes}
   \item [a] $N_{\mathrm{it}}$: average number of iterations in each {\textit{for}} loop within a time step. 
    \item [b] Algorithm \ref{prop4 implicit exp} in Appendix \ref{AppA}.
    \item [c] Algorithm \ref{prop6 implicit exp} in Appendix \ref{AppA}.
  \end{tablenotes}
\end{threeparttable}
\label{Fockbuild}
\end{table}

\subsection{Total Sampling Time}\label{SecTotT}
The last ingredient to be specified for the AutoPST approach is the total sampling time $T$, which depends not only on the desired spectral resolution but also on the chosen spectral analysis method.
The most well-known spectral analysis technique is the (discrete) Fourier transform, which
can extract all the frequencies contained in the signal at once but at the expense of a very long simulation time
($T\ge 2\pi/\delta\omega$, with $\delta\omega$ being the smallest frequency separation).
When the spectrum is block-structured (which is usually the case), it is more efficient to employ
techniques\cite{PadeFitting2016,FD1995,FD1997,FD1998,AnalyticFitting2018} that can extract a particular block of the spectrum.
Here the filter diagonalization (FD) approach\cite{FD1997} is utilized for the $K$- and $L$-edge core excitations. In essence,
FD is a nonlinear fitting process to obtain $2K$ ($\le N$) unknowns $\{d_k, \omega_k\}_{k=1}^{K}$ from $N$ data $\{C(t_n)\}_{n=0}^{N-1}$, such that
\begin{equation}
 C(t_n) = \sum_{k = 1}^K d_k e^{- i \omega_k t_n},\quad n\in[0,N-1].\label{Cfittinga}
\end{equation}
From the informational point of view, the sampling time $T$ required by FD is at least $4\pi/\delta\bar\omega$
(i.e., $2\pi/\delta\bar\omega$ for $\{d_k\}_{k=1}^{K}$ and $2\pi/\delta\bar\omega$ for $\{\omega_k\}_{k=1}^{K}$),
where $\delta\bar\omega$ is the average frequency separation in the target block of the spectrum (which is usually much larger than $\delta \omega$).
The lower bound holds when the intrinsic frequencies $\{\omega_k\}_{k=1}^{K}$ are uniformly separated.
In case that two frequencies are very close to each other, a time longer than $4\pi/\delta\bar\omega$ will be needed even if $\delta\bar\omega$ is not that small. To incorporate such features of FD, we design the following algorithm for the determination of $T$:

\begin{enumerate}[(a)]
\item For each symmetry $\Gamma$ of the excited states, determine the corresponding symmetries of the virtual ($\gamma_p$) and occupied ($\gamma_h$) CMOs and calculate their energy differences $\{\Delta \epsilon_{ai}\}|_{a\in \gamma_p, i\in \gamma_h}$.
If $|\Delta\epsilon_{m}$-$\Delta\epsilon_n|<\omega_d$ (e.g., $\omega_d=0.02$ a.u.), the
 $m$th and $n$th particle-hole pairs are considered as degenerate, such that only one of them is retained.
Sort $\{\Delta \epsilon_{ai}\}$ of all the $\gamma_p$-$\gamma_h$ pairs
in ascending order, leading to $\{\Delta\epsilon_{n}\}_{n=1}^{\Gamma,N_{\Gamma}}$.

\item Partition $\{\Delta\epsilon_{n}\}_{n=1}^{\Gamma,N_{\Gamma}}$ into blocks $\lfloor\Delta\epsilon\rfloor^{\Gamma,H_\Gamma}_{I=1}$ according to their homogeneity.
Specifically, when inserting $\Delta\epsilon_{n+1}$ into the current block renders the standard deviation of adjacent separation $\sigma_{\delta\omega}$ larger than the homogeneity $\omega_\sigma$ (e.g., $0.1$ a.u.), set $\Delta\epsilon_{n}$ as the uppermost and lowermost elements of the current and next blocks (see Fig. S1/S2 for an example). Each block $\lfloor\Delta\epsilon\rfloor^{\Gamma}_I$
is roughly uniform. The standard deviation of the mean separation, $\sigma_{\delta\bar\omega}=\sigma_{\delta\omega}/\sqrt{N_I}$ ($N_I$ is the number of intervals in the current block),
is further calculated to account for the uncertainty of $\delta\bar\omega$. This gives rise to $T_I^{\Gamma}=4\pi/(\delta\bar\omega-\sigma_{\delta\bar\omega})$.
Since the so-determined $T_I^{\Gamma}$ is merely a lower bound, it will be amplified by a factor of 1.2 for safety, thereby leading to $T_I^{\Gamma}=1.2\times 4\pi/(\delta\bar\omega-\sigma_{\delta\bar\omega})$.
\item\label{FinalT} Set $T^\Gamma=\max_I T_I^{\Gamma}$.
\end{enumerate}
As a matter of fact, there is no need to go through all the blocks in Step (\ref{FinalT}),
for the couplings between particle-hole pairs with large energetic separations are usually very small.
Instead, only the first blocks of the $K$-, $L_{2s}$-, and $L_{2p}$-edge core excitations,
partitioned by the algorithm in Sec. \ref{SecStep} with $\omega_G=0.5\ a.u.$ (see the second column in Fig. S1/S2), are to be considered.

Finally, given the time series $\{\mathbf{P}_{\beta}(t_n)\}_{n=0}^{N-1}$ ($t_n=n\Delta t$) of 1PDM, induced by an external field along a direction  $\beta\in \{x,y,z\}$,
one can calculate many interesting quantities such as the induced electric dipole moment tensor
\begin{equation}
\mu_{\alpha\beta}^{\mathrm{ind}}(t_n)=-\mathrm{Tr}[\tilde{\mathbf{V}}_{\alpha}\mathbf{P}_{\beta}(t_n)]+\mathrm{Tr}[\tilde{\mathbf{V}}_{\alpha}\mathbf{P}^{(0)}(0^-)], \quad n\in[0,N-1],\label{InduceD}
\end{equation}
which can further be transformed to the frequency domain to obtain the electric dynamical dipole polarizability
\begin{equation}
\alpha_{\alpha\beta}(\omega)=\frac{1}{\kappa}\mu_{\alpha\beta}^{\mathrm{ind}}(\omega),
\end{equation}
and hence the absorption spectrum (dipole strength function)
\begin{equation}
S(\omega)=\frac{4\pi\omega}{3c}\Im \mathrm{Tr} \boldsymbol{\alpha}(\omega).\label{DSF}
\end{equation}

\section{Pilot Applications}\label{Results}

To illustrate the usefulness of the $log_{10}(\Delta t)$-$log_{10}(|\epsilon_{1s}|)$ relations (see Table \ref{LinFit}) for XAS by CFET4/oCFET4@RT-TDDFT,
the HI and PbO molecules are taken here as showcases. The former is isovalent to the training set (HF, HCl, and HBr) but the latter is
different. The spin-free (sf) part\cite{X2CSOC1,X2CSOC2} of the exact two-component (X2C) Hamiltonian\cite{X2C2005,X2C2009},
the CAM-B3LYP\cite{CAMB3LYP} XC functional, the Dyall-TZP basis sets\cite{DyallTZP} for I and Pb, and the cc-pVTZ basis sets\cite{cc-pVTZ} for H and O are used in all calculations
with the BDF program package\cite{BDF1,BDF2,BDF3,BDFrev2020}. The interatomic distance of HI and PbO are optimized to be  1.602 {\AA} and 1.874 {\AA}, respectively, at the sf-X2C-CAM-B3LYP level.

To facilitate the dynamics simulation as well as state assignment, the full molecular symmetry should be used. For HI and PbO,
the $A_1$ irreducible representation (irrep) of $C_{\infty v}$
can be used to obtain $A_1$ type of excited states if the applied electric field is along the $z$ (bonding) direction ($z$-field for short), whereas the $E_1$ irrep of $C_{\infty v}$
should be used to obtain $E_1$ type of excited states if the applied electric field is along both the $x$ and $y$ directions.
For the latter, it is operationally much simpler to work with $C_s$, the highest subgroup of $C_{\infty v}$, where
$x$ (or $y$) transforms as the totally symmetric irrep (i.e., $C_s(\sigma_y)$ for $x$ and $C_s(\sigma_x)$ for $y$).
Since $x^2$ and $y^2$ can be rotated to each other, $\alpha_{xx}(\omega)$ is strictly identical with
$\alpha_{yy}(\omega)$, such that only one of the $x$ and $y$ directions needs to be considered
to obtain both $\alpha_{xx}(\omega)$ and $\alpha_{yy}(\omega)$. Nevertheless, only the reflection-symmetric component of
an $E_1$ state can be obtained in this case. Note in passing that it is a must instead of merely a trick to use $C_s$ if the applied $x$- or $y$-field is very strong.
In that case, both $A_1$ and $E_n$ ($n\ge2$) types of excited states can also be obtained, resulting from the
reflection-symmetric decompositions of $E_1^\beta \otimes E_1^\beta \cdots \otimes E_1^\beta$,
$\beta\in\{x, y\}$. In contrast, the $A_2$ type of excited states still cannot be obtained, unless the angular momentum $l_z$ is used in lieu of the electric dipole moment.
Having understood the symmetry issues (i.e., the particle-hole pairs must belong to the same irreps for both $z$-field under $C_{\infty v}$ and $x/y$-field under $C_s(\sigma_{y/x})$),
the time steps $\Delta t$ and total simulation times $T$
can be determined according to the algorithms described in Secs. \ref{SecStep} and \ref{SecTotT}, respectively, see Table \ref{TimeStep and T}.
It is of interest to see that the so-determined time steps for oCFET4/CFET4 are larger/smaller than the Nyquist ones
($6.24\times 10^{-5}$ fs for HI and $2.35\times 10^{-5}$ fs for PbO).

\begin{table}[h]
\centering
\caption{Predicted time steps ($\Delta t$) and total simulation times ($T$) of CFET4/oCFET4@RT-TDDFT/CAM-B3LYP for core excitations of HI and PbO}
\small
\begin{threeparttable}
\begin{tabular}{cccccccc}\toprule
molecule & \makecell{field\\direction} & \makecell{$\Delta t$/fs\\ CFET4} & \makecell{$\Delta t$/fs\\ oCFET4} & $T$/fs \\
\midrule
\multirow{2}{*}{HI}
            & z & $4.38\times 10^{-5}$ & $9.26\times 10^{-5}$ & $4.06$ \\
            & x & $4.38\times 10^{-5}$ & $9.26\times 10^{-5}$ & $5.28$ \\
\multirow{2}{*}{PbO}
            & z & $1.66\times 10^{-5}$ & $3.51\times 10^{-5}$ & $6.79$ \\
            & x & $1.66\times 10^{-5}$ & $3.51\times 10^{-5}$ & $7.05$ \\
\bottomrule
\end{tabular}
\end{threeparttable}
\label{TimeStep and T}
\end{table}

\begin{figure}[h]
\centering
\begin{tabular}{ccc}
\includegraphics[width=0.45\textwidth]{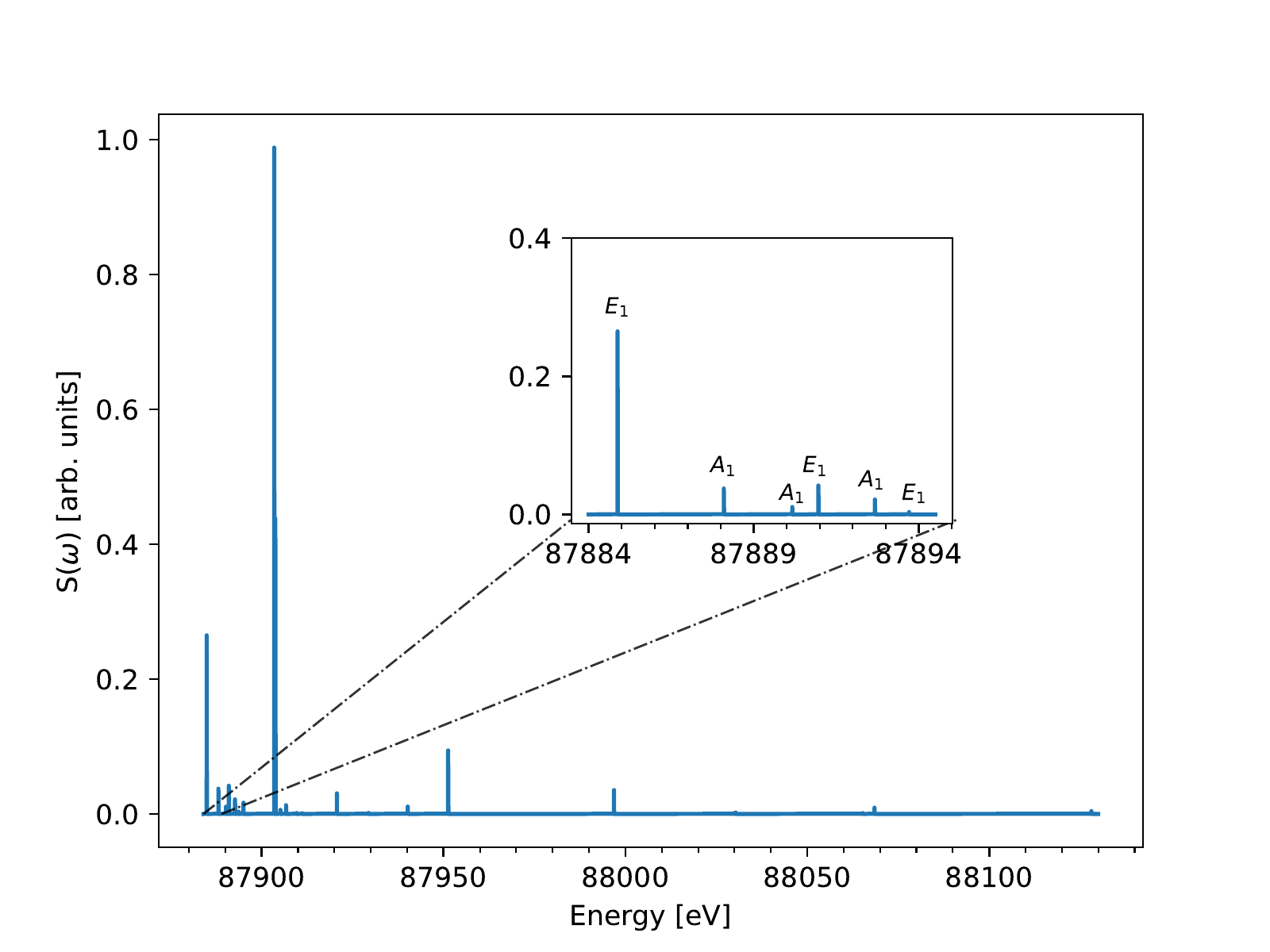}\\
(a)\\
\includegraphics[width=0.45\textwidth]{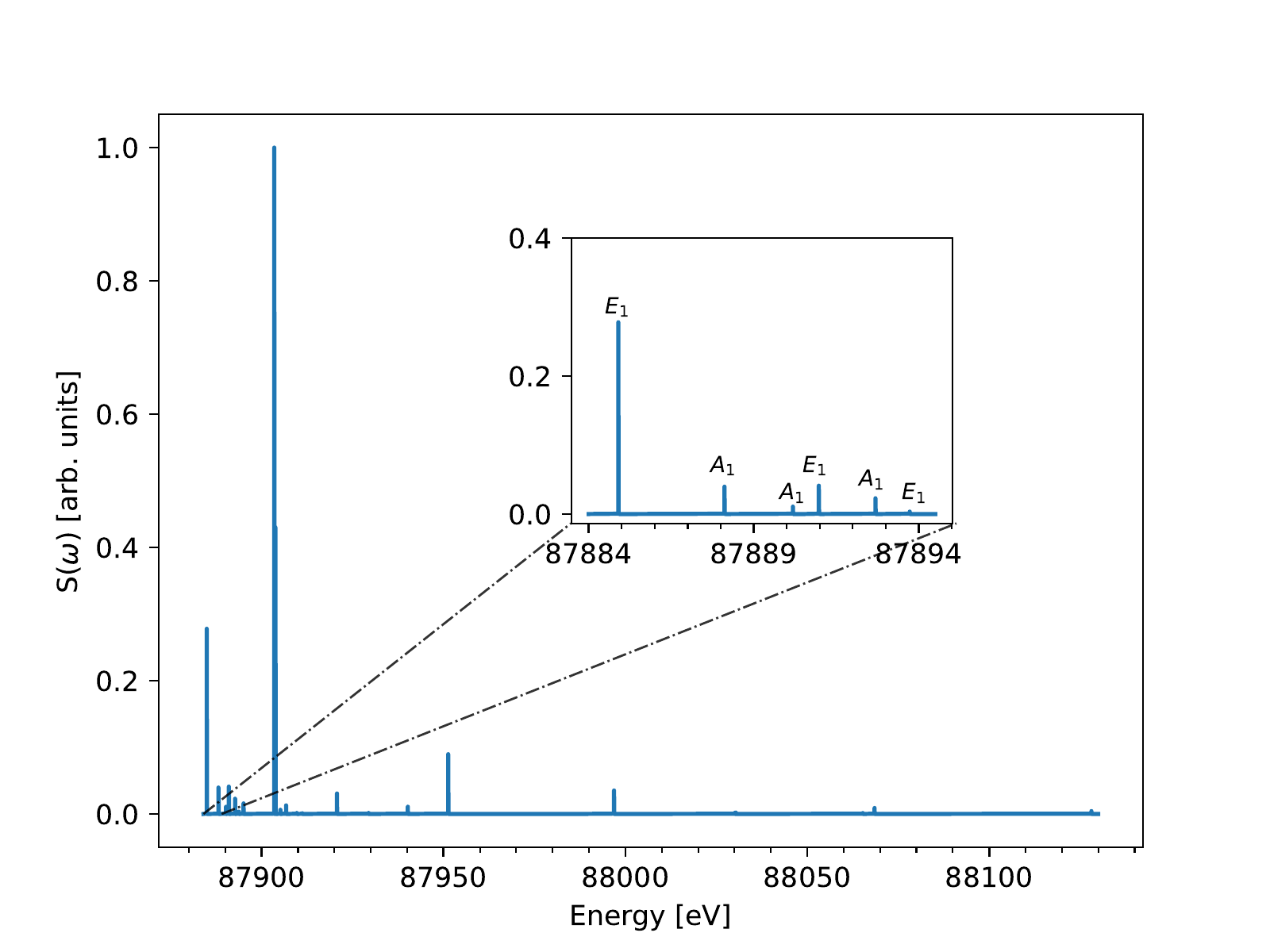}\\
(b)\\
\includegraphics[width=0.45\textwidth]{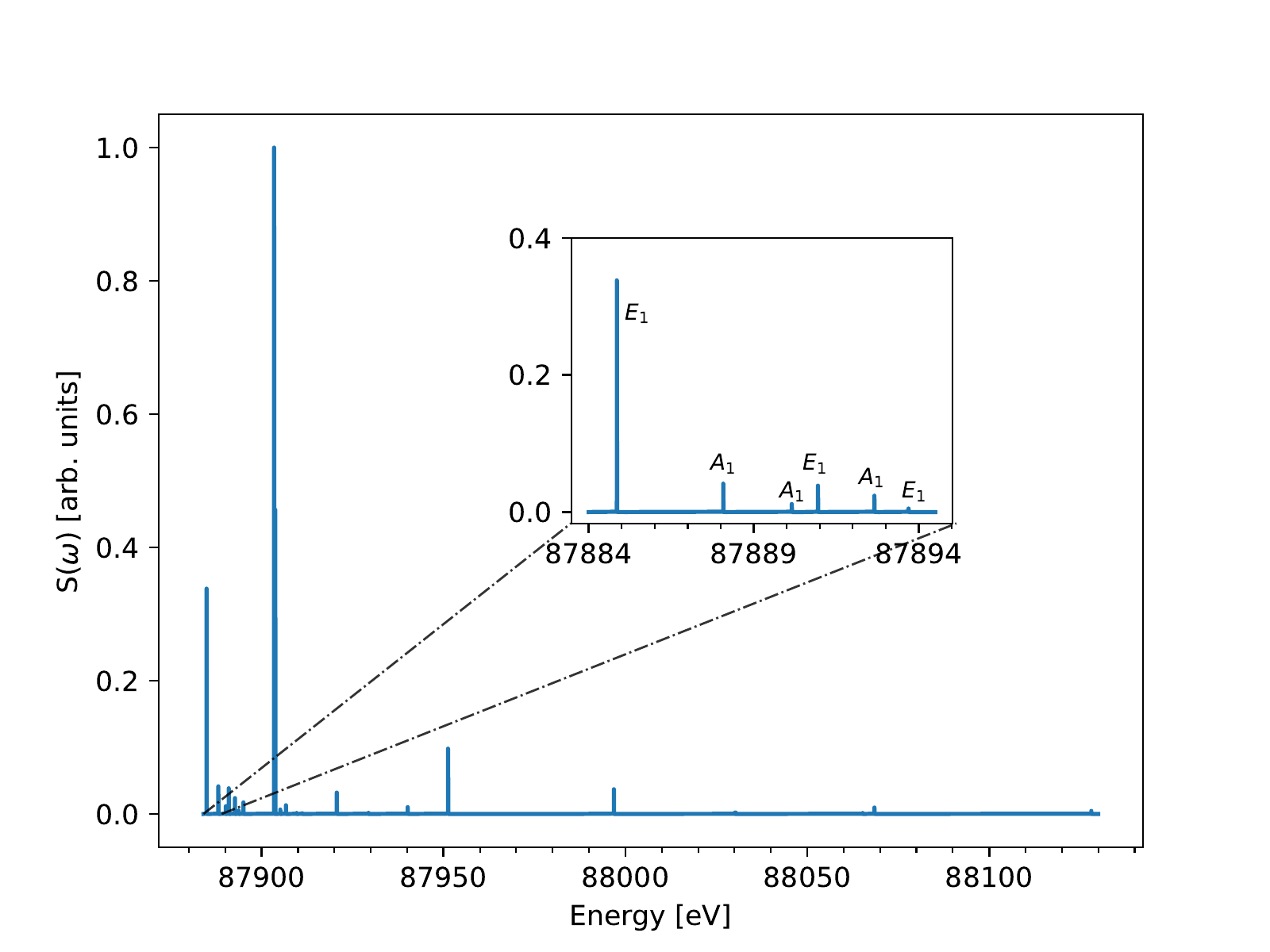}\\
(c)\\
\end{tabular}
\caption{$K$-edge absorption spectra of PbO simulated by (a) CFET4@RT-TDDFT/CAM-B3LYP, (b) oCFET4@RT-TDDFT/CAM-B3LYP, and (c) LR-TDDFT/CAM-B3LYP.}\label{PbO Kedge fig}
\end{figure}

\begin{figure}[h]
\centering
\begin{tabular}{ccc}
\includegraphics[width=0.45\textwidth]{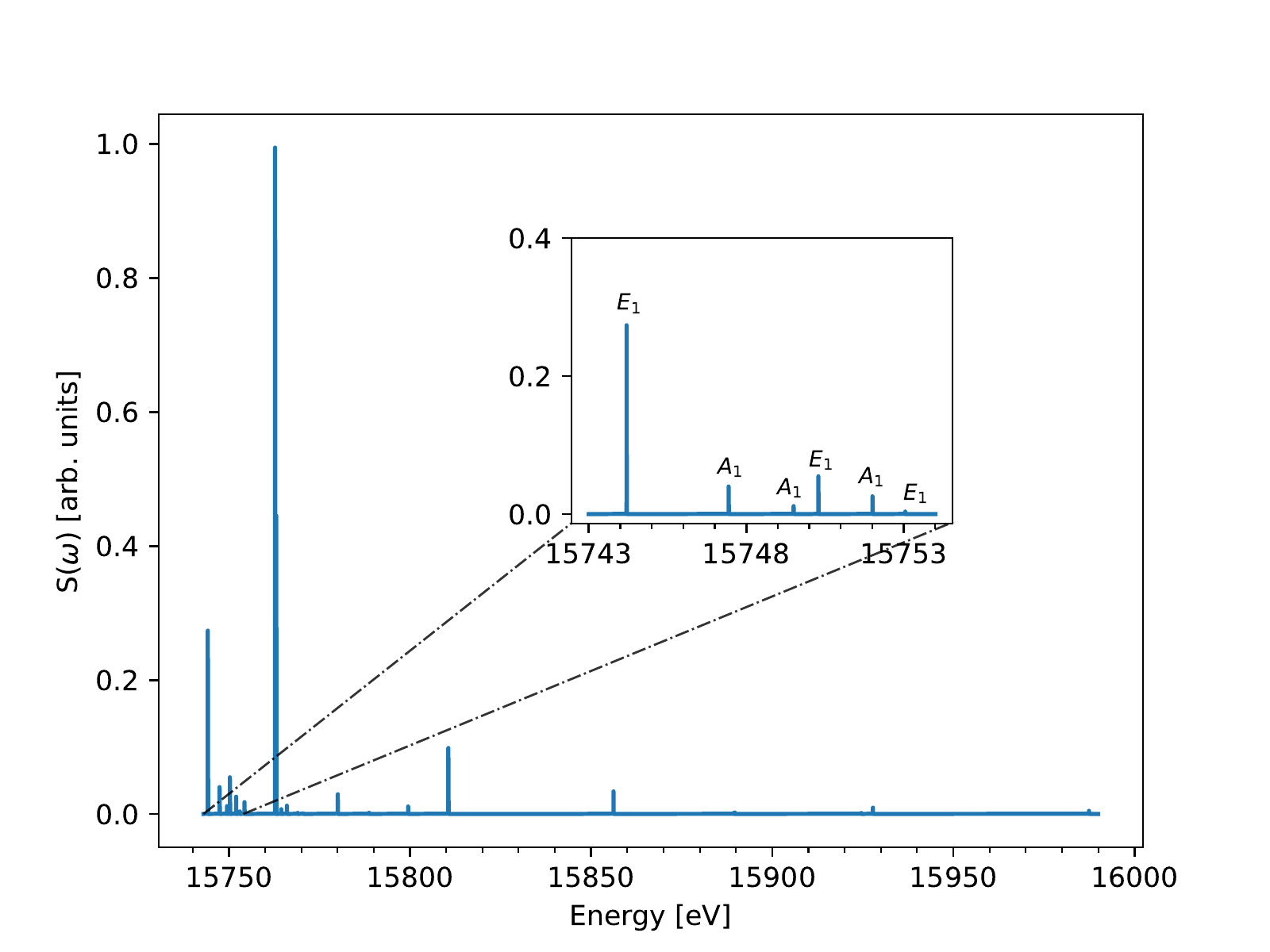}\\
(a)\\
\includegraphics[width=0.45\textwidth]{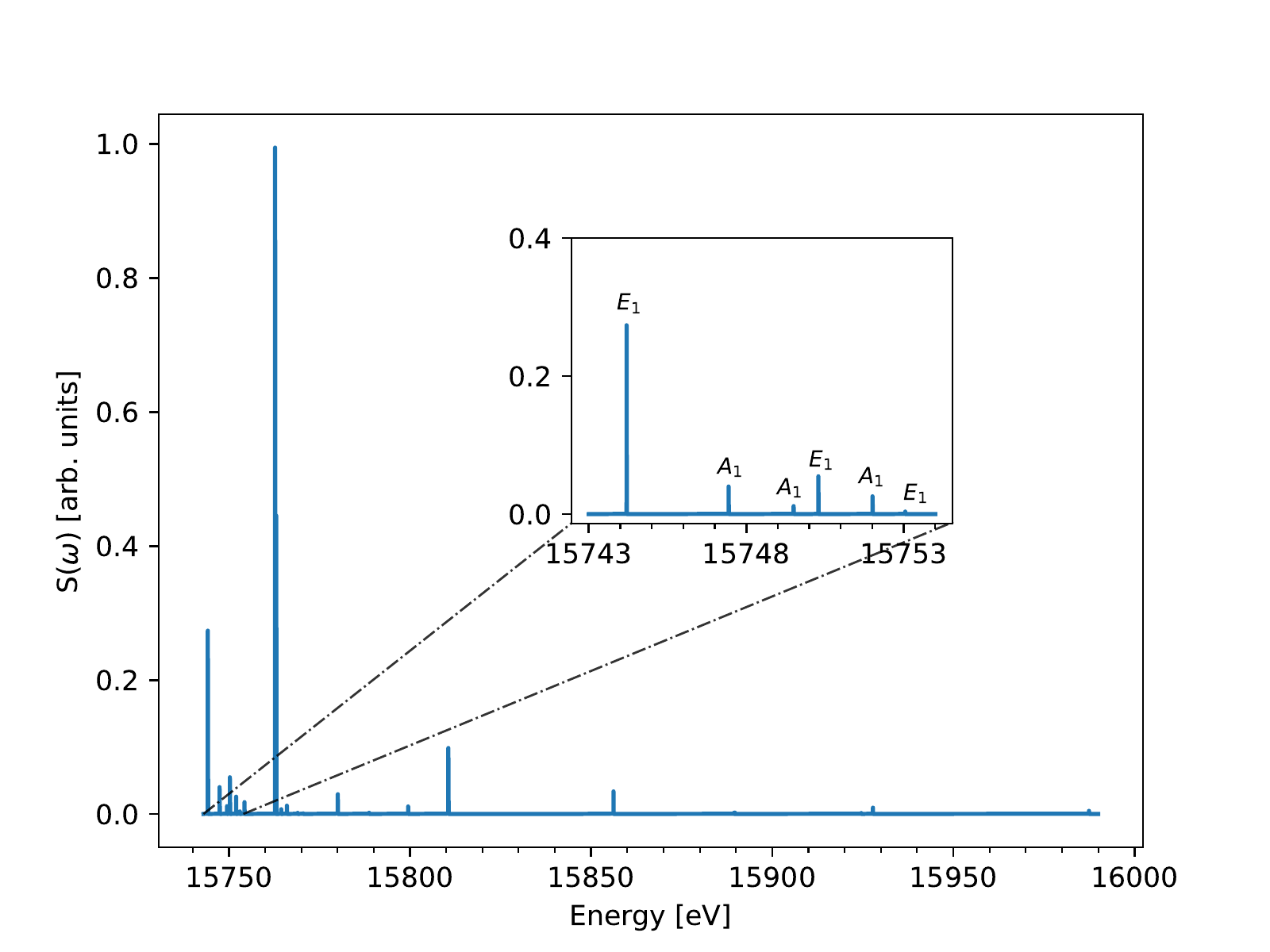}\\
(b)\\
\includegraphics[width=0.45\textwidth]{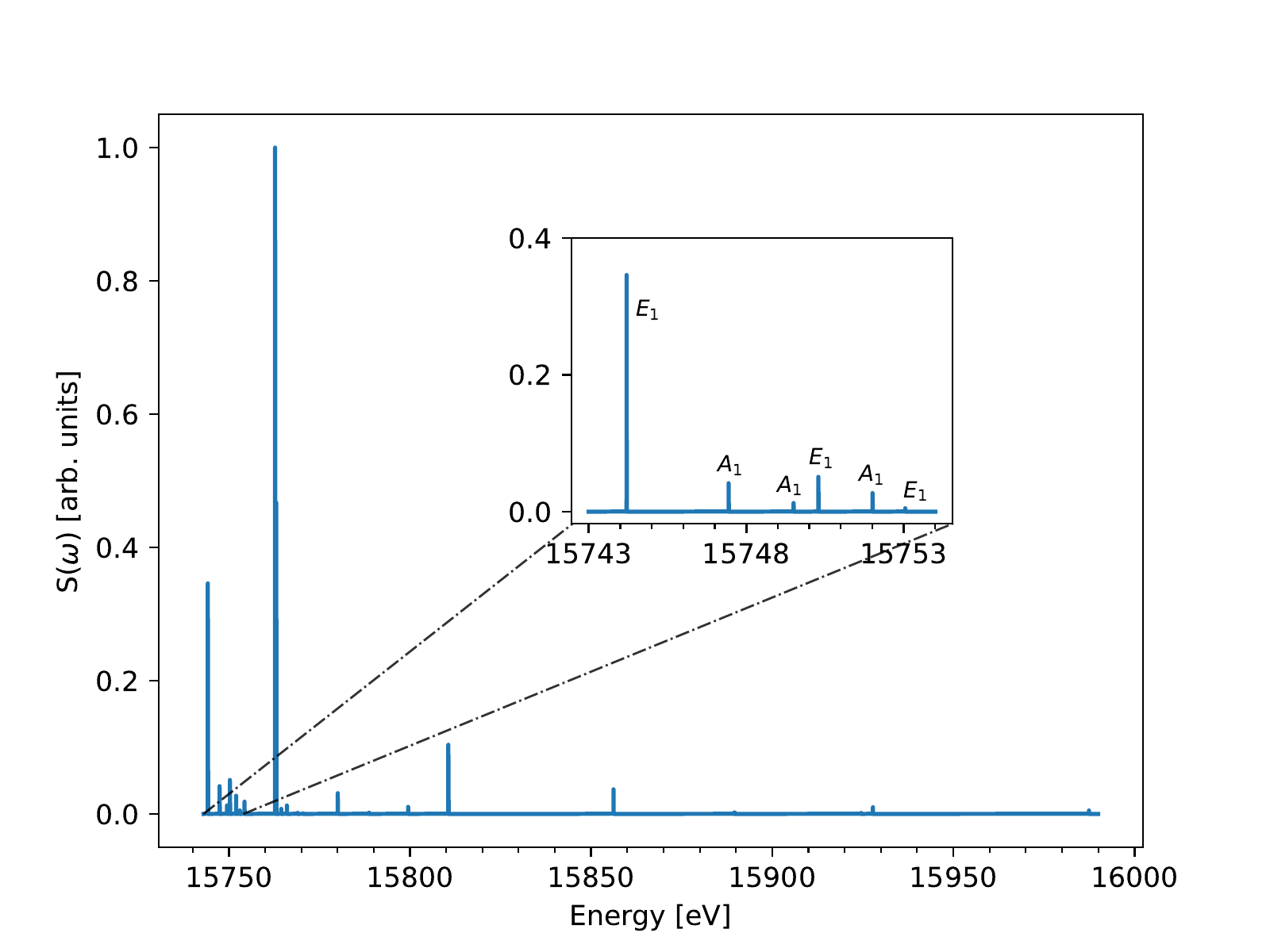}\\
(c)\\
\end{tabular}
\caption{$L_{2s}$-edge absorption spectra of PbO simulated by (a) CFET4@RT-TDDFT/CAM-B3LYP, (b) oCFET4@RT-TDDFT/CAM-B3LYP, and (c) LR-TDDFT/CAM-B3LYP.}\label{PbO Ledge2s fig}
\end{figure}

\begin{figure}[h]
\centering
\begin{tabular}{ccc}
\includegraphics[width=0.45\textwidth]{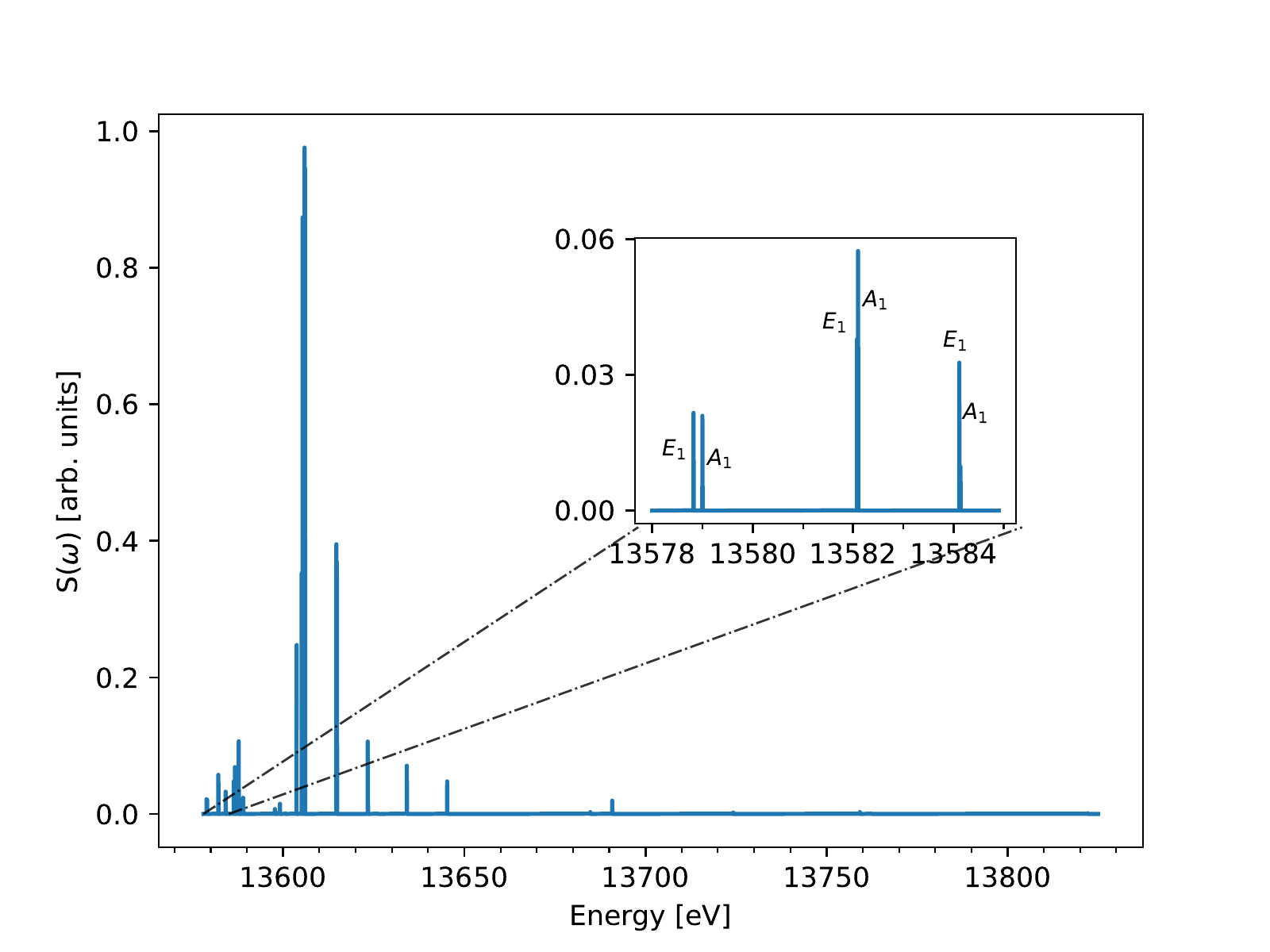}\\
(a)\\
\includegraphics[width=0.45\textwidth]{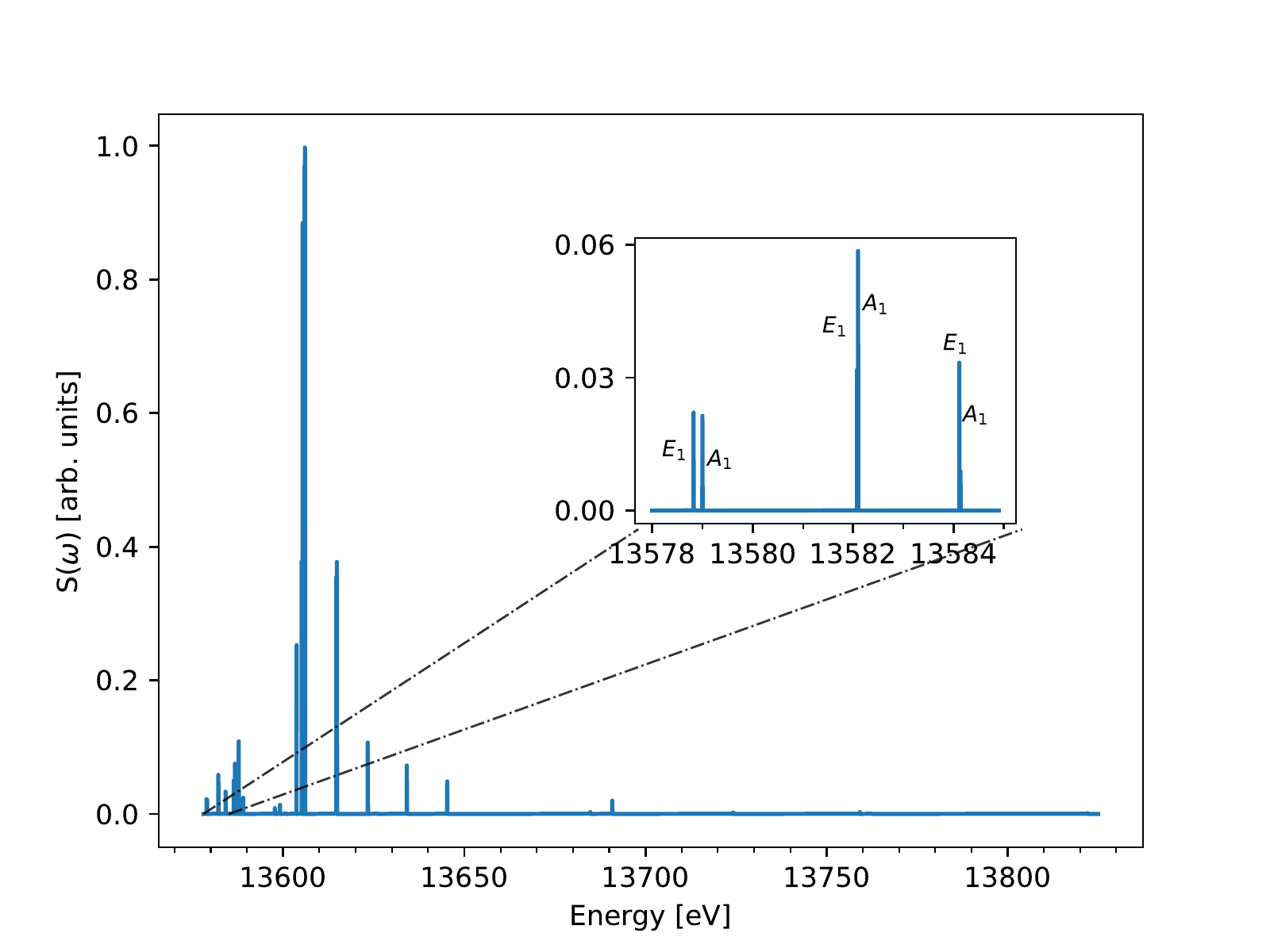}\\
(b)\\
\includegraphics[width=0.45\textwidth]{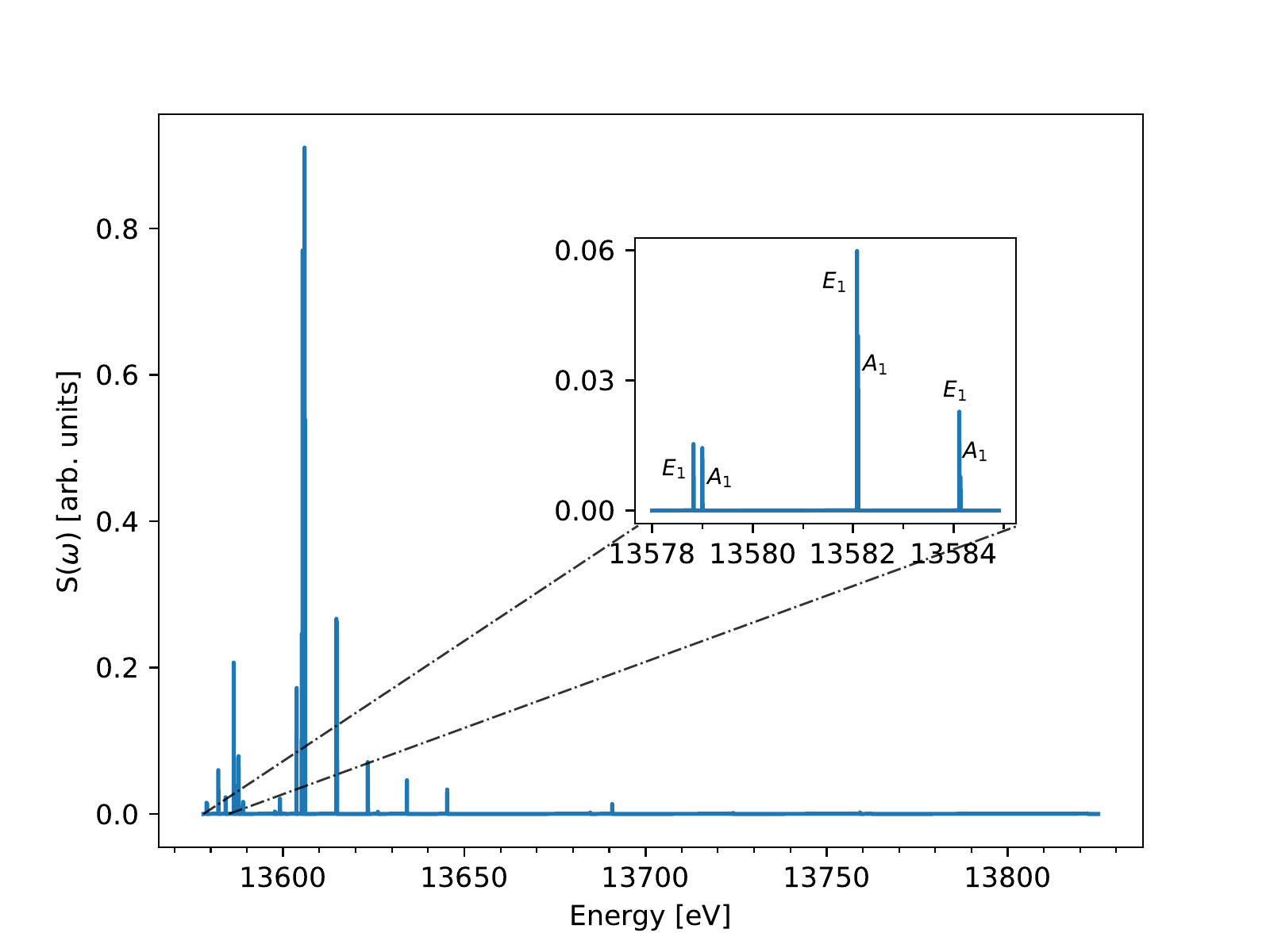}\\
(c)\\
\end{tabular}
\caption{$L_{2p}$-edge absorption spectra of PbO simulated by (a) CFET4@RT-TDDFT/CAM-B3LYP, (b) oCFET4@RT-TDDFT/CAM-B3LYP, and (c) LR-TDDFT/CAM-B3LYP.}\label{PbO Ledge2p fig}
\end{figure}

\begin{figure}[h]
\centering
\begin{tabular}{ccc}
\includegraphics[width=0.5\textwidth]{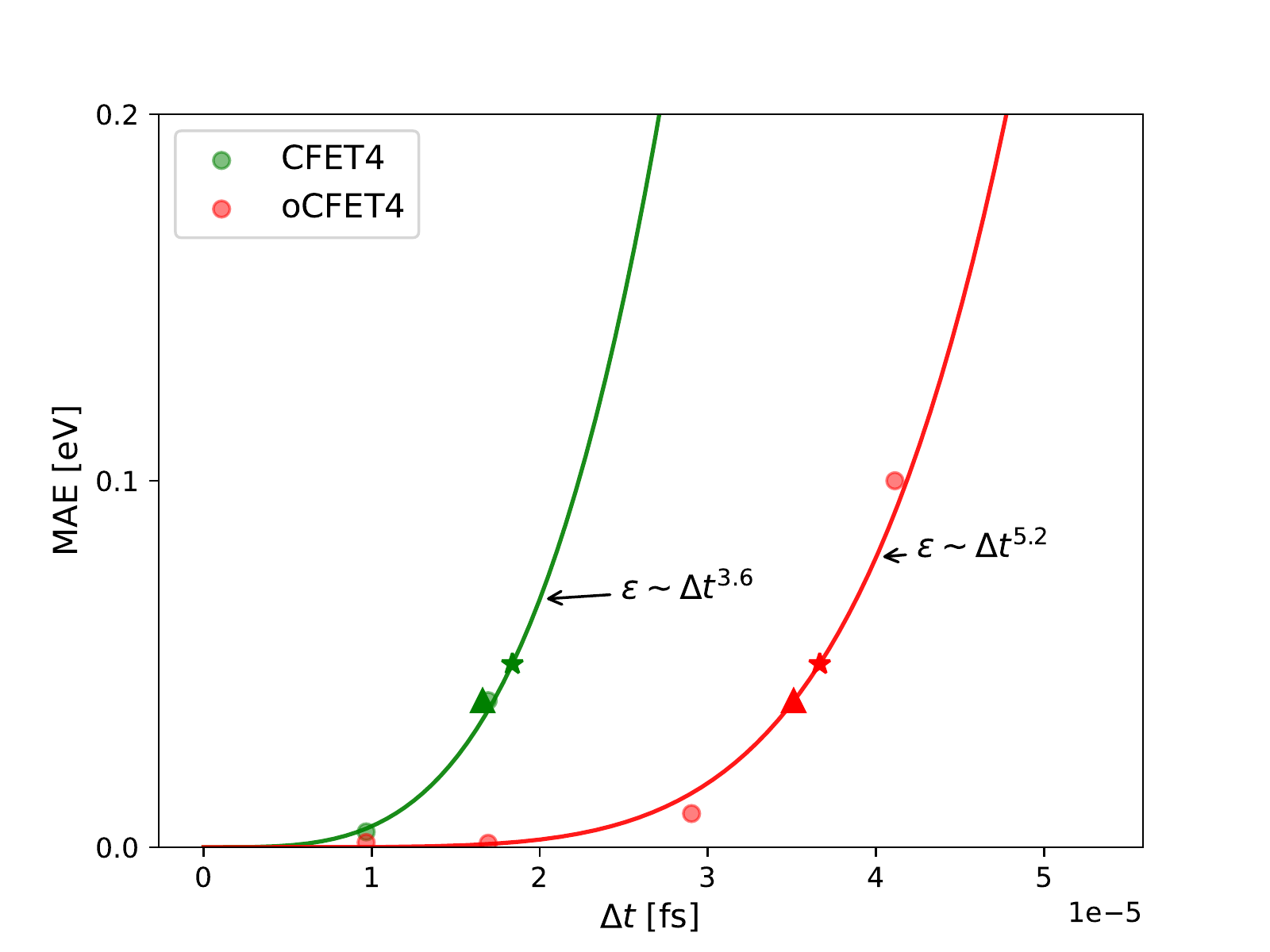}\\
(a)\\
\includegraphics[width=0.5\textwidth]{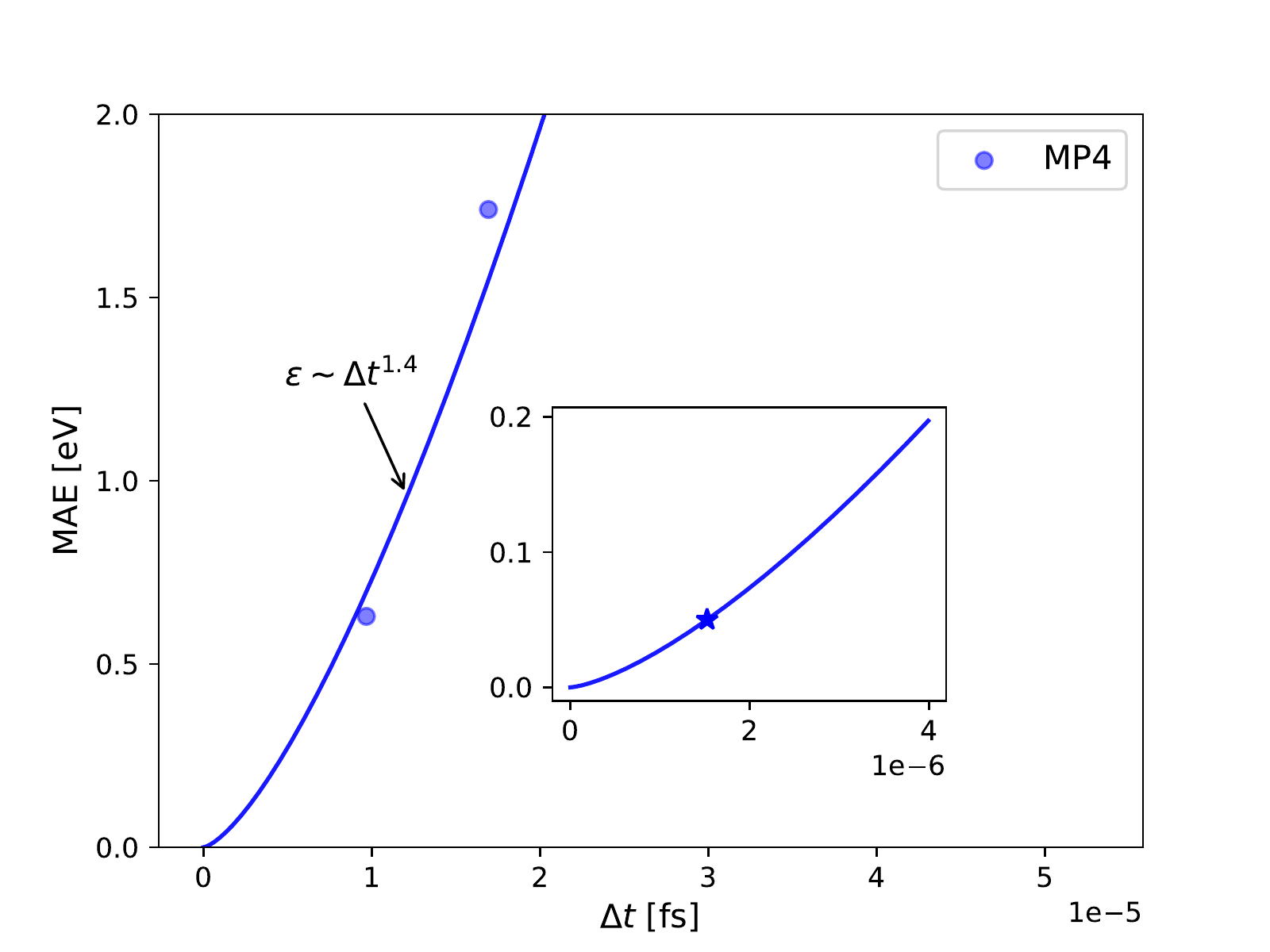}\\
(b)\\
\end{tabular}
\caption{Mean absolute error (MAE) of (a) CFET4/oCFET4@RT-TDDFT/CAM-B3LYP and (b) MP4@RT-TDDFT/CAM-B3LYP as function of time steps for $K$-edge excitations of PbO under the z-field.
The optimal time steps for MAE $= 0.05$ eV are marked as asterisks, while the
predicted time steps (cf. Table \ref{TimeStep and T}) are marked as triangles. }\label{Error_dt}
\end{figure}

The $K$- and $L$-edge absorption spectra of PbO (HI) can be obtained by the FD analysis\cite{FD1997} of the CFET4@RT-TDDFT/CAM-B3LYP and oCFET4@RT-TDDFT/CAM-B3LYP signals for 
each 1PDM element $P_{ai}(t)$ (more precisely $P_{ai}(t)+P_{ia}(t)=2\Re P_{ai}(t)$). As can be seen from 
 Figs. \ref{PbO Kedge fig}-\ref{PbO Ledge2p fig} (Figs. S3-S5), the so-obtained spectra are in excellent agreement with those by LR-TDDFT/CAM-B3LYP,
where the fully symmetrized\cite{Symm2009} iterative vector interaction (iVI) approach\cite{iVI,iVI-TDDFT} is employed to obtain directly the core excitations as interior roots of LR-TDDFT,
without knowing in advance the number or characters of the roots.  
To make the comparison even sharper, the $K$-, $L_{2s}$-, and $L_{2p}$-edge core
excitation energies of PbO and HI are further given in Tables S5-S7 and Tables S8-S10, respectively.
It can be seen that the deviations of CFET4/oCFET4@RT-TDDFT from iVI-LR-TDDFT\cite{iVI,iVI-TDDFT} are all below 0.05 eV, with the mean absolute error being 0.04 eV for both PbO and HI.
This is just the tolerance when training the $log_{10}(\Delta t)$-$log_{10}(|\epsilon_{1s}|)$ linear relations with respect to the lowest five bright
$K$-edge core excited states of HX (X = F, Cl, Br). Moreover, it can be seen from Fig. \ref{Error_dt} that
the predicted time steps are indeed nearly optimal. As such, the robustness of these relations is fully confirmed.

Having obtained the absorption spectra, it remains to see how individual particle-hole pairs contribute to the spectra.
Instead of using\cite{RT4CTD2015} the imaginary parts of
the isotropically averaged induced dipole moment pairs $\bar{\mu}_{ai}^{\mathrm{ind}}(\omega_I)$
\begin{align}
\bar{\mu}_{ai}^{\mathrm{ind}}(\omega_I)&=\frac{1}{3}\sum_{\beta\in\{x,y,z\}}\tilde{V}^{(1)}_{\beta,ia}d_{\beta,ai}(\omega_I),\label{AvMu}\\
d_{\beta,ai}(\omega_I)&= \int_{-\infty}^{+\infty}  [P_{\beta,ai}(t)+P_{\beta,ia}(t)]e^{i\omega_I t}dt,\label{FTPai}
\end{align}
the imaginary parts of $\{d_{\beta,ai}(\omega_I)\}$ are employed here to estimate the compositions $T_{ai}(\omega_I)$ of state $I$:
\begin{align}
T_{ai}(\omega_I)&=\frac{ \bar{P}_{ai}^2(\omega_I)} { \sum_{bj}\bar{P}_{bj}^2(\omega_I)},\\
\bar{P}_{ai}^2(\omega_I)&=\frac{1}{3}\sum_{\beta\in\{x,y,z\}} [\Im d_{\beta,ai}(\omega_I)]^2.\label{Composition}
\end{align}
Noticing that $d_{\beta,ai}(\omega_I)$ \eqref{FTPai} is just the (complex) amplitude (associated with $\omega_I$)
in the expression \eqref{Cfittinga} for the signal $2\Re P_{ai}(t)$ of each particle-hole pair,
it can readily be obtained from the FD analysis\cite{FD1997}, which usually requires a much shorter simulation time than
the Fourier transform \eqref{FTPai}.
 As can be seen from Tables S5-S10, the so-obtained
transition compositions are indeed very close to those according to the eigenvectors of iVI-LR-TDDFT.
Had $\Im \bar{\mu}_{ai}^{\mathrm{ind}}(\omega_I)$ \eqref{AvMu} been used,
the transition compositions would be scaled down by the electric dipole moment matrix elements $\tilde{V}^{(1)}_{ai}$, so as to deteriorate
the agreement. As a matter of fact, it can be seen from equation (104) in Ref. \citenum{lwjTDRev} that $P_{ai}(\omega_I)$ and $[P_{ai}(-\omega_I)]^*$
 ($\equiv P_{ai}^*(\omega_I)$) are just proportional, respectively, to the upper ($X_{ai,I}$) and lower ($Y_{ai,I}$) components of the $I$th eigenvector of LR-TDDFT,
 i.e., $P_{ai}(\omega_I)\sim i X_{ai,I} (\mathbf{X}_I^\dag \tilde{\mathbf{V}}^{(1)})$ and $P_{ai}^*(\omega_I)\sim i Y_{ai,I} (\mathbf{Y}_I^\dag \tilde{\mathbf{V}}^{(1)})$. Therefore, $\Im d_{\beta,ai}(\omega_I)\sim X_{ai,I} (\mathbf{X}_I^\dag \tilde{\mathbf{V}}^{(1)})
 + Y_{ai,I} (\mathbf{Y}_I^\dag \tilde{\mathbf{V}}^{(1)})$ is related directly to the coefficient $\{C_{ai,I}\}$ of the excited state $|I\rangle$ in the basis of singly excited determinants
(i.e., $|I\rangle=\sum_{ai}C_{ai,I}a_a^\dag a_i|0\rangle$; cf. equation (113) in Ref. \citenum{lwjTDRev}).
 
\begin{figure}[h]
\centering
\begin{tabular}{cccc}
\includegraphics[width=0.45\textwidth]{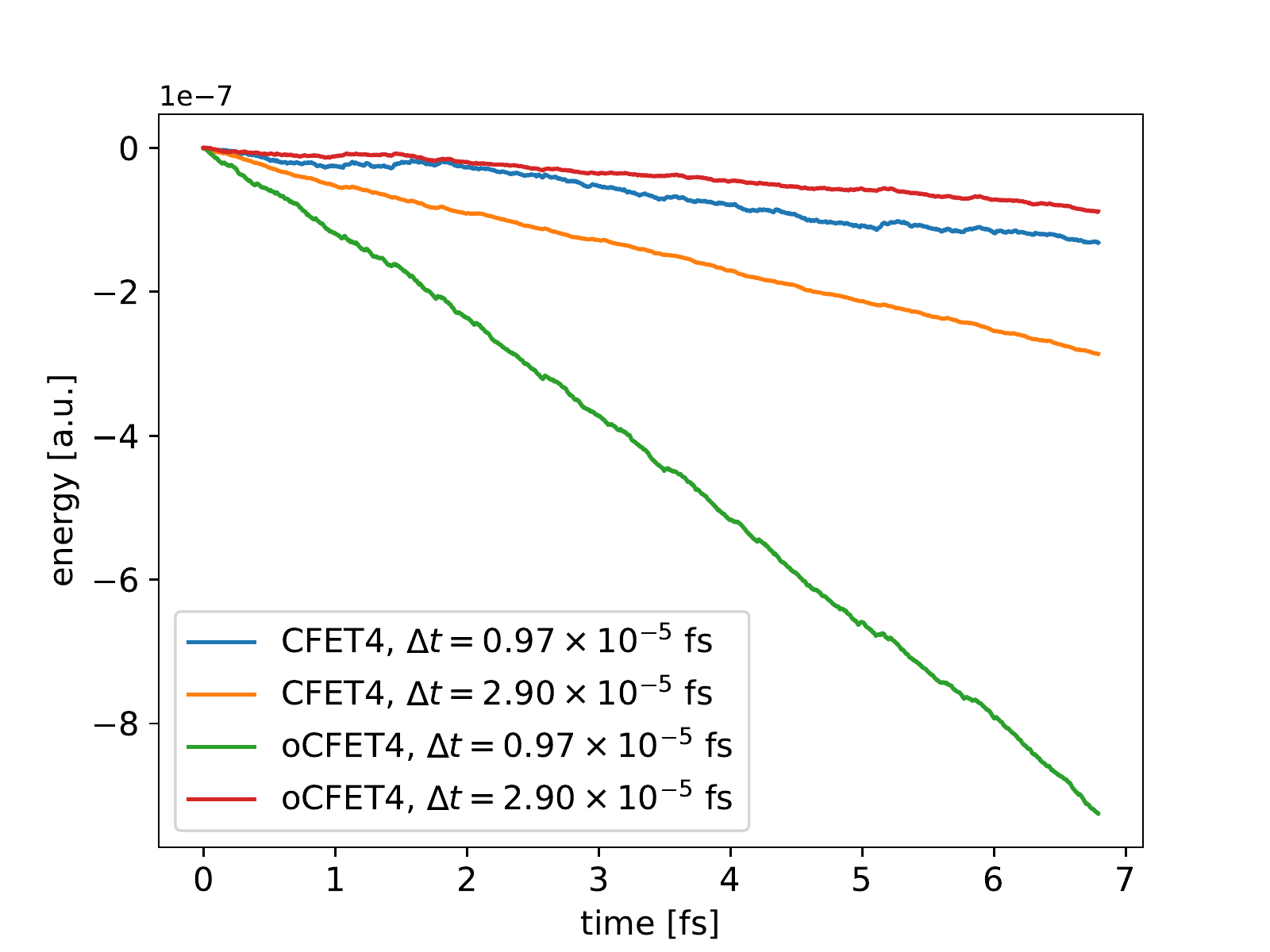}\\
(a)\\
\includegraphics[width=0.45\textwidth]{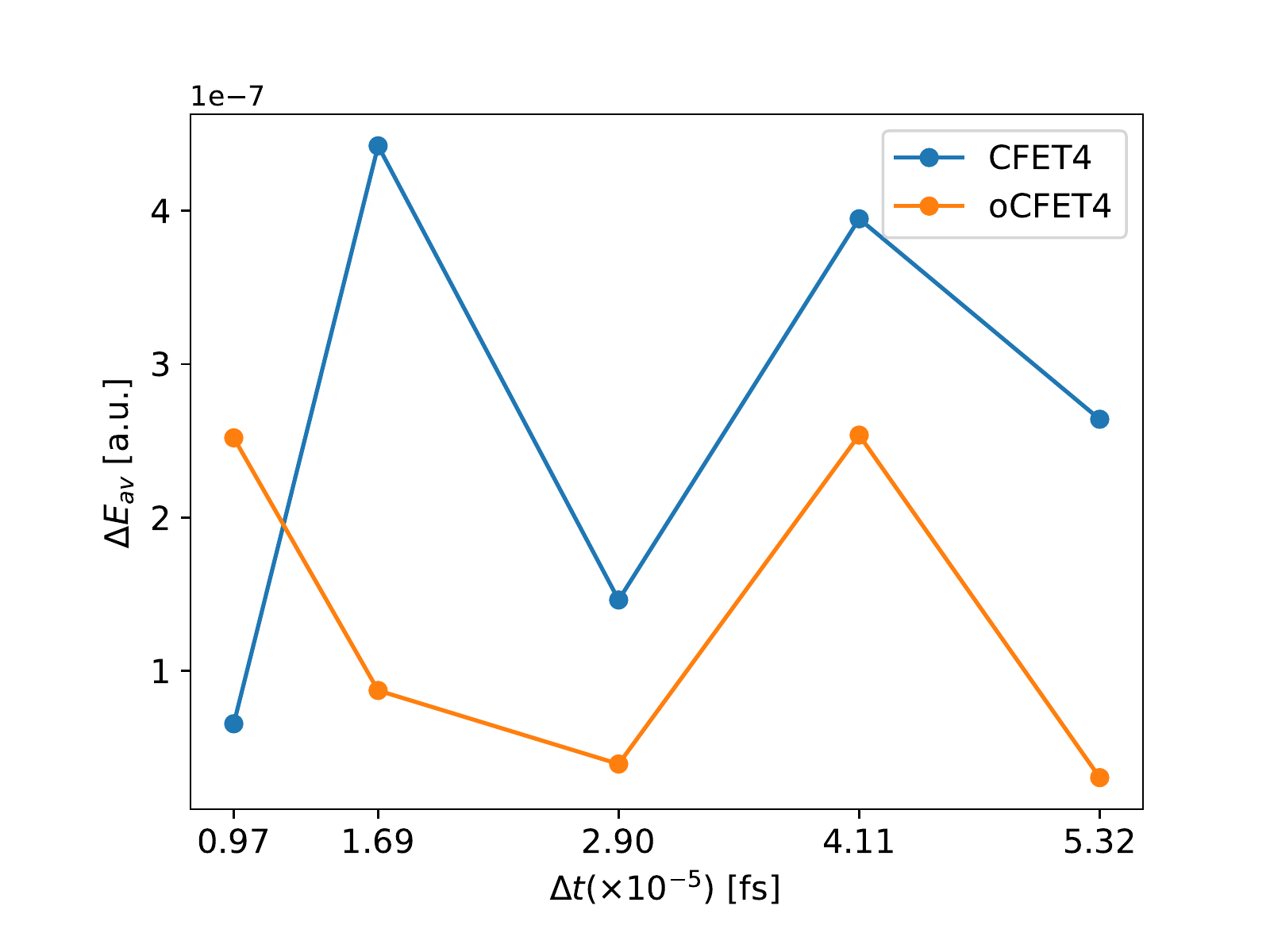}\\
(b)\\
\end{tabular}
\caption{(a) Energy conservation error $\Delta E_k(\Delta t)=E(t_k)-E(0^+)$ as function of simulation time and (b)
average energy conservation error $\Delta E_{av}(\Delta t)=\frac{1}{N}\sum_{k=1}^N |\Delta E_k(\Delta t)|$ as function of time steps.
CFET4/oCFET4@RT-TDDFT/CAM-B3LYP results for PbO under the z-field.}
\label{PbO-fluctuationofE}
\end{figure}

Some remarks on the stability and efficiency of the algorithms are now in order. For the former,
it can be seen from Fig. \ref{PbO-fluctuationofE}(a) that
the energy conservation errors (ECE), $\Delta E_k(\Delta t)=E(t_k)-E(0^+)$, are indeed
very small for both CFET4@RT-TDDFT/CAM-B3LYP and oCFET4@RT-TDDFT/CAM-B3LYP, although they grow roughly linearly for each time step
as a result of error accumulation.
It can also be seen from Fig. \ref{PbO-fluctuationofE}(b) that the ECEs, quantified by
$\Delta E_{av}(\Delta t)=\frac{1}{N}\sum_{k=1}^N|\Delta E_k(\Delta t)|$, are not monotonic with respect to time step: a larger time step may lead to a smaller ECE, which is also observed in other propagtors\cite{PropagatorsKS2018}. It has been attempted to
modify the EPEP2 (EPEP3) predictor-corrector by putting $\mathbf{P}(t_1^{[4]})$ ($\mathbf{P}(t_1^{[6]})$) also into the iteration cycle, by backward Euler or ETRS propagation from $\mathbf{P}(t_2^{[4]})$ ($\mathbf{P}(t_2^{[6]})$). However, the stability is not improved discernibly.
Anyhow, such small ECEs do not affect the spectral accuracy even when the total simulation times are doubled (see Table S11).

\begin{figure}[h]
\centering
\begin{tabular}{ccc}
\includegraphics[width=0.5\textwidth]{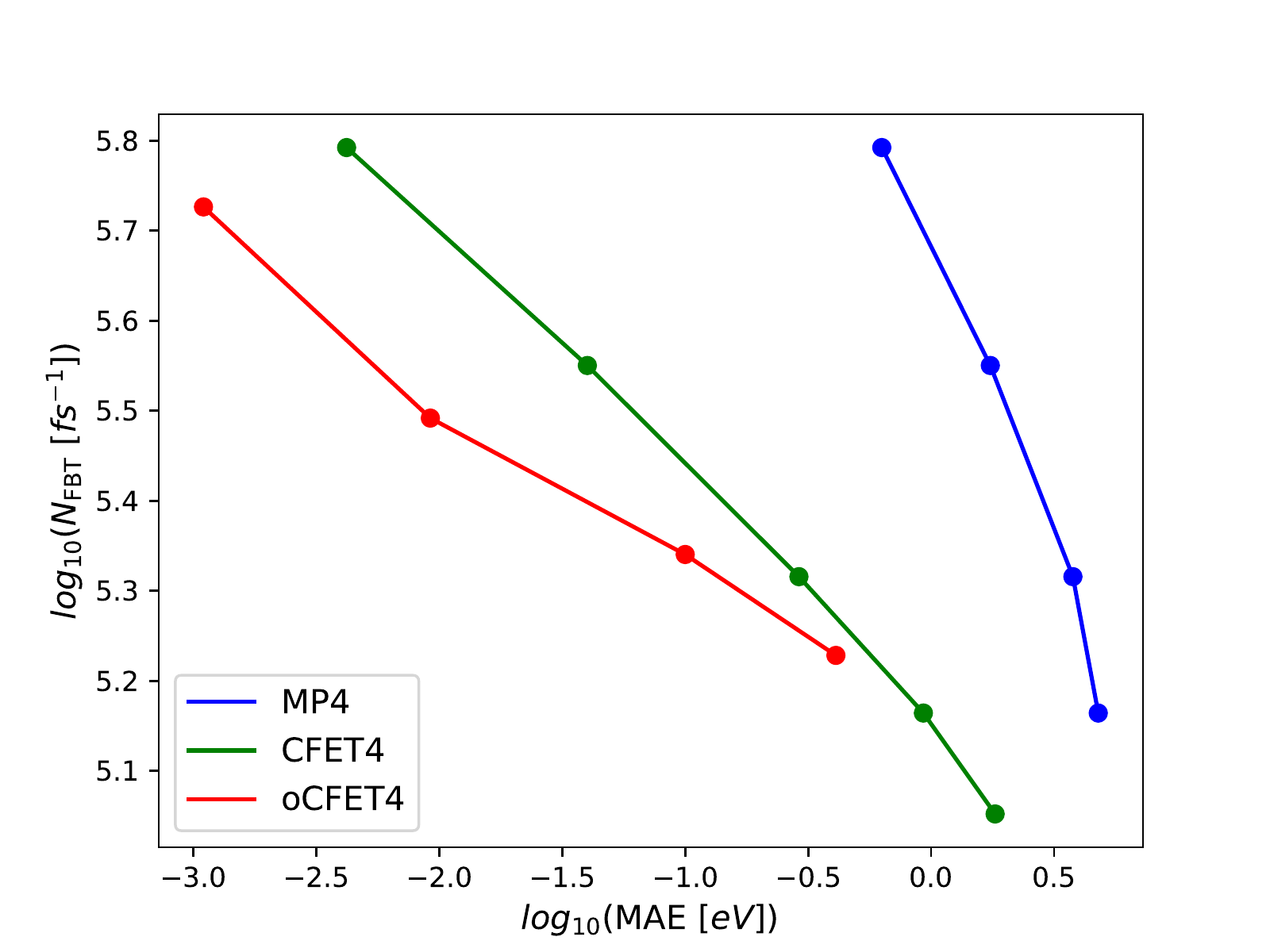}\\
\end{tabular}
\caption{Number of Fock builds per unit time ($N_{\mathrm{FBT}}$) as function of the mean absolute error (MAE)
by MP4/CFET4/oCFET4@RT-TDDFT/CAM-B3LYP for $K$-edge excitations of PbO under the z-field (cf. Table S12).}\label{FPT_error}
\end{figure}

Since the same total simulation time is used, the relative computational cost of CFET4@RT-TDDFT and oCFET4@RT-TDDFT
can be estimated by the ratio between their $N_{\mathrm{FBS}}/\Delta t$ values (cf. Tables \ref{Fockbuild} and \ref{TimeStep and T}).
In the present simulations, both EPEP2 in CFET4 and EPEP3 in oCFET4 require two cycles to converge at each time step, such that
the value of $N_{\mathrm{FBS}}/\Delta t$ is equivalent to the number of Fock builds per unit time ($N_{\mathrm{FBT}}$).
It can then readily be checked that CFET4@RT-TDDFT is ca. 1.4 times as expensive as oCFET4@RT-TDDFT for both HI and PbO.
That is, the computational overhead of oCFET4@RT-TDDFT over CFET4@RT-TDDFT per time step is compensated for by its larger time step
and hence fewer number of steps. As can be seen from Fig. \ref{FPT_error}, oCFET4@RT-TDDFT remains to be more efficient than CFET4@RT-TDDFT
for a wide range of spectroscopic accuracy. In contrast, MP4@RT-TDDFT is much more expensive.
As a matter of fact, MP4@RT-TDDFT is not competitive even for systems containing only light atoms.
For instance, it is ca. 2.2 times as expensive as oCFET4@RT-TDDFT to achieve an accuracy of $<0.05$ eV for the $K$-edge excitations of HF.
Admittedly, even the most efficient scheme considered here (i.e.,
oCFET4@RT-TDDFT) is more than four orders of magnitude more expensive than iVI-LR-TDDFT\cite{iVI,iVI-TDDFT} (see Table S13), solely
due to the necessity of using a very small time step to fully resolve the core excitation spectra. However,
the situation may be changed for wide absorption spectra of large systems with high densities of states, as demonstrated before\cite{RT-LR2015}.


\section{Conclusion}\label{Conclusion}
A very robust AutoPST approach has been proposed to run relativistic RT-TDDFT simulations of XAS of systems composed of any elements in the Periodic Table.
It is composed of the following ingredients: (1) the just good enough and most efficient propagator oCFET4\cite{oCFET2011} along with the EPEP3 predictor-corrector, (2)
very robust linear relations and block aliasing conditions for determining automatically the largest possible time steps,
(3) \emph{a priori} estimates of the total simulation times, and (4) the use of full molecular symmetry.
Although only scalar relativity has been taken into account in the pilot applications,
AutoPST should work well also for fully relativistic RT-TDDFT, even beyond the linear response regime.

\begin{acknowledgments}
This work was supported by the Key-Area Research and Development Program of Guangdong Province (Grant No. 2020B0101350001),
National Natural Science Foundation of China (Grant Nos. 21833001 and 21973054), and Mountain Tai Climbing Program of Shandong Province.
\end{acknowledgments}

\section*{Data Availability Statement}
The data that support the findings of this study are available within the article and its supplementary material.

\section*{Conflicts of interest}
There are no conflicts to declare.

\section*{Supporting Information}
Orbital energies, core excitation energies, and absorption spectra of HI and PbO.

\clearpage
\newpage

\appendix

\section{Algorithms for CFET4 and oCFET4}\label{AppA}

\begin{algorithm}[H]
\caption{Fourth-order commutator-free exponential time-propagator (CFET4) with two exponentials
   (equation (43) in Ref.\citenum{CFET2006})}\label{mag42}
\begin{flushleft}
\textbf{Subroutine} CFET4($\mathbf{P}(0)$, $\mathbf{F}(t_1)$, $\mathbf{F}(t_2)$)
\end{flushleft}
\begin{algorithmic}[1]
\Inputs {$\mathbf{P}(0)$, $\mathbf{F}(t_1)$, $\mathbf{F}(t_2)\gets t_1=(\frac{1}{2}-\frac{\sqrt{3}}{6})\Delta t$, $t_2=(\frac{1}{2}+\frac{\sqrt{3}}{6})\Delta t$}
\State $\mathbf{U}_1(\Delta t,0)\gets e^{-i\left(\frac{3+2\sqrt{3}}{12}\mathbf{F}(t_1)+\frac{3-2\sqrt{3}}{12}\mathbf{F}(t_2)\right)\Delta t}$
\State $\mathbf{P}(\Delta t)\gets\mathbf{U}_1(\Delta t,0)\mathbf{P}(0)\mathbf{U}_1^\dagger(\Delta t,0)$
\State $\mathbf{U}_2(\Delta t,0)\gets e^{-i\left(\frac{3-2\sqrt{3}}{12}\mathbf{F}(t_1)+\frac{3+2\sqrt{3}}{12}\mathbf{F}(t_2)\right)\Delta t}$
\State $\mathbf{P}(\Delta t)\gets \mathbf{U}_2(\Delta t,0)\mathbf{P}(\Delta t)\mathbf{U}_2^\dagger(\Delta t,0)$
\Outputs {$\mathbf{P}(\Delta t)$}
\end{algorithmic}
\end{algorithm}

\begin{algorithm}[H]
\caption{Optimized fourth-order commutator-free exponential time-propagator (oCFET4) with three exponentials
  [see equation (43) in Ref.\citenum{oCFET2011}. The coefficients $x_{ik}$ in Eq. \eqref{fijdef} are determined by expressing
the coefficients $\mathbf{A}_n$ defined in equation (25) in Ref. \citenum{oCFET2011} in terms of the matrix moments $\mathbf{A}_j^{[2s]}$ \eqref{Aijdef}]}\label{mag43}
\begin{flushleft}
\textbf{Subroutine} oCFET4($\mathbf{P}(0), \mathbf{F}(t_1), \mathbf{F}(t_2), \mathbf{F}(t_3)$)
\end{flushleft}
\begin{algorithmic}[1]
\Inputs {$\mathbf{P}(0)$\\
$\mathbf{F}(t_1)\gets t_1=(\frac{1}{2}-\frac{\sqrt{15}}{10})\Delta t$\\
$\mathbf{F}(t_2)\gets t_2=\frac{1}{2}\Delta t$\\
$\mathbf{F}(t_3)\gets t_3=(\frac{1}{2}+\frac{\sqrt{15}}{10})\Delta t$}
\State $a \leftarrow \left(\begin{array}{ccc}
      \frac{37}{240} - \frac{10 \sqrt{15}}{261} & - \frac{1}{30} &
      \frac{37}{240} + \frac{10 \sqrt{15}}{261}\\
      - \frac{11}{360} & \frac{23}{45} & - \frac{11}{360}\\
      \frac{37}{240} + \frac{10 \sqrt{15}}{261} & - \frac{1}{30} &
      \frac{37}{240} - \frac{10 \sqrt{15}}{261}
    \end{array}\right)$
\State $\mathbf{U}_1 (\Delta t, 0)\gets e^{- i\Delta t (a_{31}\mathbf{F} (t_1) + a_{32} \mathbf{F} (t_2) + a_{33}\mathbf{F} (t_3))}$
\State $\mathbf{P}(\Delta t)\gets\mathbf{U}_1 (\Delta t, 0) \mathbf{P} (0)\mathbf{U}_1^{\dagger} (\Delta t, 0)$
\State $\mathbf{U}_2(\Delta t,0)\gets e^{- i\Delta t(a_{21}\mathbf{F}(t_1)+a_{22}\mathbf{F}(t_2)+a_{23}\mathbf{F} (t_3))}$
\State $\mathbf{P} (\Delta t)\gets\mathbf{U}_2(\Delta t,0)\mathbf{P}(\Delta t)\mathbf{U}_2^{\dagger}(\Delta t, 0)$
\State $\mathbf{U}_3(\Delta t,0)\gets e^{- i\Delta t (a_{11}\mathbf{F}(t_1)+a_{12}\mathbf{F}(t_2)+a_{13}\mathbf{F}(t_3))}$
\State $\mathbf{P}(\Delta t)\gets\mathbf{U}_3(\Delta t,0)\mathbf{P}(\Delta t)\mathbf{U}_3^\dagger (\Delta t,0)$
\Outputs {$\mathbf{P}(\Delta t)$}
\end{algorithmic}
\end{algorithm}

\begin{algorithm}[H]
\caption{Exponential prediction of density matrix and exponential correction of density matrix at two Gauss-Legendre points (EPEP2) for CFET4}\label{prop4 implicit exp}
\begin{algorithmic}[1]
\Inputs {$\mathbf{P}(0^-) \gets$ ground state DFT density matrix \\
$t_{0}, t_{f}, \Delta t$\\
$\tilde{\mathbf{V}}^{(1)} \gets$ initial perturbation}
\Initialize {$N \gets (t_{f}-t_{0})/\Delta t$\\
$\mathbf{U}(0^+,0^-) \gets e^{-i\tilde{\mathbf{V}}^{(1)}}$\\
$\mathbf{P}(0)\equiv\mathbf{P}(0^+)\gets \mathbf{U}(0^+,0^-)\mathbf{P}(0^-)\mathbf{U}^\dagger(0^+,0^-)$}
\For{step $k=0$ to $N-1$}\Comment{Propagation loop}
\State $t_k\gets k\Delta t$
\State $t_a\gets t_k+(\frac{1}{2}-\frac{\sqrt{3}}{6})\Delta t$
\State $t_b\gets t_k+(\frac{1}{2}+\frac{\sqrt{3}}{6})\Delta t$
\State $\mathbf{P}_1(t_a)\gets \mathrm{ETRS\_1}(\mathbf{P}(t_k),t_k,t_a)$ \Comment{Step 1 (Algorithm \ref{etrs1})}
\State $\mathbf{F}_1(t_a)\gets \mathrm{Fock}(\mathbf{P}_1(t_a))$\Comment{Step 2}
\State $\mathbf{U}_1(t_b,t_a)\gets e^{-i\mathbf{F}_1(t_a)(t_b-t_a)}$
\State $\mathbf{P}_1(t_b)\gets\mathbf{U}_1(t_b,t_a)\mathbf{P}_1(t_a)\mathbf{U}_1^\dagger(t_b,t_a)$\Comment{Step 3}
\State $\mathbf{F}_1(t_b)\gets \mathrm{Fock}(\mathbf{P}_1(t_b))$\Comment{Step 4}
\State $\mathbf{P}_2(t_{k+1})\gets\mathrm{CFET4}(\mathbf{P}(t_k),\mathbf{F}_1(t_a),\mathbf{F}_1(t_b))$\Comment{Step 5 (Algorithm \ref{mag42})}

\For{iteration $i=2$ to $N_{\mathrm{it}}$} \Comment{Loop until $\Vert\mathbf{P}_{i+1}(t_{k+1})-\mathbf{P}_i(t_{k+1})\Vert_F<M\xi$}
\State $\mathbf{F}_i(t_{k+1})\gets\mathrm{Fock}(\mathbf{P}_i(t_{k+1}))$\Comment{Step 6}
\State $\mathbf{U}_i(t_b,t_{k+1})\gets e^{-i\mathbf{F}_i(t_{k+1})(t_{k+1}-t_b)}$
\State $\mathbf{P}_i(t_b)\gets\mathbf{U}_i(t_b,t_{k+1})\mathbf{P}_i(t_{k+1})\mathbf{U}_i^\dagger(t_b,t_{k+1})$\Comment{Step 7}
\State $\mathbf{F}_i(t_b)\gets\mathrm{Fock}(\mathbf{P}_i(t_b))$\Comment{Step 4}
\State $\mathbf{P}_{i+1}(t_{k+1})\gets\mathrm{CFET4}(\mathbf{P}(t_k),\mathbf{F}_1(t_a),\mathbf{F}_i(t_b))$\Comment{Step 5 (Algorithm \ref{mag42})}
\EndFor
\State $\mathbf{P}(t_{k+1})\gets\mathbf{P}_{N_{\mathrm{it}}+1}(t_{k+1})$
\EndFor
\end{algorithmic}
\end{algorithm}

\begin{algorithm}[H]
\caption{One-step enforced time reversal symmetry propagator (ETRS, Eq. \eqref{etrsprop})}
\label{etrs1}
\begin{flushleft}
\textbf{Subroutine} ETRS\_1($\mathbf{P}(t_0)$, $t_0$, $t_1$)
\end{flushleft}
\begin{algorithmic}[1]
\Inputs {$\mathbf{P}(t_0)$, $t_0$, $t_1$}
\State $\Delta t\gets t_1 - t_0$
\State $\mathbf{F}(t_0)\gets \mathrm{Fock}(\mathbf{P}(t_0))$
\State $\mathbf{U}_0(t_0+\frac{\Delta t}{2},t_0)\gets e^{-i\mathbf{F}(t_0)\frac{\Delta t}{2}}$
\State $\mathbf{P}_1(t_0+\frac{\Delta t}{2})\gets \mathbf{U}_0(t_0+\frac{\Delta t}{2},t_0)\mathbf{P}(t_0)\mathbf{U}_0^\dagger(t_0+\frac{\Delta t}{2},t_0)$ \Comment{Step 1}
\State $\mathbf{U}_1(t_1,t_0)\gets e^{-i\mathbf{F}_1(t_0)\Delta t}$
\State $\mathbf{P}_2(t_1)\gets \mathbf{U}_1(t_1,t_0)\mathbf{P}(t_0)\mathbf{U}_1^\dagger(t_1,t_0)$ \Comment{Step 2}
\For{iteration $i=2$ to $N_{\mathrm{it}}$}\Comment{Loop until $\Vert\mathbf{P}_{i+1}(t_{k+1})-\mathbf{P}_i(t_{k+1})\Vert_F<M\xi$}
\State $\mathbf{F}_i(t_1)\gets \mathrm{Fock}(\mathbf{P}_i(t_1))$\Comment{Step 3}
\State $\mathbf{U}_i(t_1,t_0+\frac{\Delta t}{2})\gets e^{-i\mathbf{F}_i(t_1)\frac{\Delta t}{2}}$
\State $\mathbf{P}_{i+1}(t_1)\gets \mathbf{U}_i(t_1,t_0+\frac{\Delta t}{2})\mathbf{P}(t_0+\frac{\Delta t}{2})\mathbf{U}_i^\dagger(t_1,t_0+\frac{\Delta t}{2})$\Comment{Step 4}
\EndFor
\Outputs {$\mathbf{P}_{N_{\mathrm{it}}+1}(t_1)$}
\end{algorithmic}
\end{algorithm}

\begin{algorithm}[H]
\scriptsize
\caption{Exponential prediction of density matrix and exponential correction of density matrix at three Gauss-Legendre points (EPEP3) for the oCFET4 propagator}\label{prop6 implicit exp}
\begin{algorithmic}[1]
\Inputs {$\mathbf{P}(0^-) \gets$ ground state DFT density matrix \\
$t_{0}, t_{f}, \Delta t$\\
$\tilde{\mathbf{V}}^{(1)} \gets$ initial perturbation}
\Initialize {$N \gets (t_{f}-t_{0})/\Delta t$\\
$\mathbf{U}(0^+,0^-) \gets e^{-i\tilde{\mathbf{V}}^{(1)}}$\\
$\mathbf{P}(0)\equiv\mathbf{P}(0^+)\gets \mathbf{U}(0^+,0^-)\mathbf{P}(0^-)\mathbf{U}^\dagger(0^+,0^-)$}
\For{step $k=0$ to $N-1$}\Comment{Propagation loop}
\State $t_k\gets k\Delta t$
\State $t_a\gets t_k+(\frac{1}{2}-\frac{\sqrt{15}}{10})\Delta t$
\State $t_b\gets t_k+\frac{1}{2}\Delta t$
\State $t_c\gets t_k+(\frac{1}{2}+\frac{\sqrt{15}}{10})\Delta t$
\State $t_d\gets\frac{t_b+t_c}{2}$
\State $\mathbf{P}_1(t_a)\gets\mathrm{ETRS\_1}(\mathbf{P}(t_k),t_k,t_a)$\Comment{Step 1 (Algorithm \ref{etrs1})}
\State $\mathbf{F}_1(t_a)\gets\mathrm{Fock}(\mathbf{P}_1(t_a))$\Comment{Step 2}
\State $\mathbf{U}_1(t_b,t_a)\gets e^{-i\mathbf{F}_1(t_a)(t_b-t_a)}$
\State $\mathbf{P}_1(t_b)\gets\mathbf{U}_1(t_b,t_a)\mathbf{P}_1(t_a)\mathbf{U}_1^\dagger(t_b,t_a)$\Comment{Step 3}
\State $\mathbf{F}_1(t_b)\gets\mathrm{Fock}(\mathbf{P}_1(t_b))$\Comment{Step 4}
\State $\mathbf{U}_1(t_c,t_a)\gets e^{-i\mathbf{F}_1(t_b)(t_c-t_a)}$
\State $\mathbf{P}_1(t_c)\gets \mathbf{U}_1(t_c,t_a)\mathbf{P}_1(t_a)\mathbf{U}_1^\dagger(t_c,t_a)$\Comment{Step 5}
\State $\mathbf{F}_1(t_c)\gets\mathrm{Fock}(\mathbf{P}_1(t_c))$\Comment{Step 6}
\State $\mathbf{P}_2(t_{k+1})\gets\mathrm{oCFET4}(\mathbf{P}(t_k),\mathbf{F}_1(t_a),\mathbf{F}_1(t_b),\mathbf{F}_1(t_c))$\Comment{Step 7 (Algorithm \ref{mag43})}
\For{iteration $i=2$ to $N_{\mathrm{it}}$}\Comment{Loop until $\Vert\mathbf{P}_{i+1}(t_{k+1})-\mathbf{P}_i(t_{k+1})\Vert_F<M\xi$}
\State $\mathbf{F}_i(t_{k+1})\gets \mathrm{Fock}(\mathbf{P}_i(t_{k+1}))$\Comment{Step 8}
\State $\mathbf{U}_i(t_c,t_{k+1})\gets e^{i\mathbf{F}_i(t_{k+1})(t_{k+1}-t_c)}$
\State $\mathbf{P}_i(t_c)\gets \mathbf{U}_i(t_c,t_{k+1})\mathbf{P}_i(t_{k+1})\mathbf{U}_i^\dagger(t_c,t_{k+1})$\Comment{Step 9}
\State $\mathbf{F}_i(t_c)\gets \mathrm{Fock}(\mathbf{P}_i(t_c))$\Comment{Step 10}
\State $\mathbf{U}_i(t_d,t_c)\gets e^{i\mathbf{F}_i(t_c)\frac{t_c-t_b}{2}}$
\State $\mathbf{P}_i(t_d)\gets\mathbf{U}_i(t_d,t_c)\mathbf{P}_i(t_c)\mathbf{U}_i^\dagger(t_d,t_c)$\Comment{Step 11}
\State $\mathbf{F}_i(t_d)\gets\mathrm{Fock}(\mathbf{P}_i(t_d))$\Comment{Step 12}
\State $\mathbf{U}_i(t_b,t_c)\gets e^{i\mathbf{F}_i(t_d)(t_c-t_b)}$
\State $\mathbf{P}_i(t_b)\gets\mathbf{U}_i(t_b,t_c)\mathbf{P}_i(t_c)\mathbf{U}_i^\dagger(t_b,t-c)$\Comment{Step 13}
\State $\mathbf{F}_i(t_b)\gets\mathrm{Fock}(\mathbf{P}_i(t_b))$\Comment{Step 14}
\State $\mathbf{P}_{i+1}(t_{k+1})\gets\mathrm{oCFET4}(\mathbf{P}(t_k),\mathbf{F}_1(t_a),\mathbf{F}_i(t_b),\mathbf{F}_i(t_c))$\Comment{Step 7 (Algorithm \ref{mag43})}
\EndFor
\State $\mathbf{P} (t_{k + 1}) \gets\mathbf{P}_{N_{\mathrm{it}} + 1} (t_{k + 1})$
\EndFor
\end{algorithmic}
\end{algorithm}

\clearpage
\newpage

\bibliographystyle{apsrev4-1}
\bibliography{rtTDDFT}

\end{document}